\documentstyle[epsf]{mnsty}
\def \farcs{\hbox{$.\!\!^{\prime\prime}$}}
\def \farcm{\hbox{$.\!\!^{\prime}$}}
\def \moverl{\hbox{${\rm M}_\odot/{\rm L}_{{\rm B}\odot}$}}
\def \lumstar{{${\rm L}^*_{\rm B}(z=0)=5.6\times10^9~h^{-2} {\rm L}_{\rm {B}\odot}~$}}

\def \lstar{{${\rm L}^*_{\rm B}$}} 

\title[Lensing by galaxies in CNOC2 fields]
{Lensing by galaxies in CNOC2 fields$^\star$}

\author [Hoekstra et al.]
{H.~Hoekstra$^{1,2,3}$, M.~Franx$^4$, K.~Kuijken$^3$, 
R.G.~Carlberg$^{2,5}$, H.K.C.~Yee$^{2,5}$\\
	$^1$~CITA, University of Toronto, 60 St. George Street,
	Toronto, M5S 3H8, Canada\\
	$^2$~Department of Astronomy, University of Toronto, 
	60 St. George Street, Toronto, M5S 3H8, Canada\\
	$^3$~Kapteyn Astronomical Institute, University of Groningen, 
        Postbus 800, 9700 AV Groningen, The Netherlands \\
	$^4$~Leiden Observatory, P.O.~Box 9513, 2300 RA Leiden, 
	The Netherlands\\
	$^5$~Visiting Astronomer, Canada-France-Hawaii Telescope}

\begin{document}
 
\maketitle

\begin{abstract}
We have observed two blank fields of approximately 30 by 23 arcminutes
using the William Herschel Telescope. The fields have been
studied as part of the Canadian Network for Observational Cosmology
Field Galaxy Redshift Survey (CNOC2), and spectroscopic
redshifts are available for 1125 galaxies in the two fields.  
We measured the lensing signal caused by large scale structure, and 
found that the result is consistent with current, more accurate 
measurements.

We study the galaxy-galaxy lensing signal of three overlapping samples
of lenses (one with and two without redshift information), and detect
a significant signal in all cases. The estimates for the velocity
dispersion of an \lumstar\ galaxy agree well for the various
samples.  The best fit singular isothermal sphere model to the
ensemble averaged tangential distortion around the galaxies with
redshifts yields a velocity dispersion of $\sigma_*=130^{+15}_{-17}$
km/s, or a circular velocity of $V_c^*=184^{+22}_{-25}$ km/s for an
\lstar\ galaxy, in good agreement with other studies.

We use a maximum likelihood analysis, where a parameterized mass model
is compared to the data, to study the extent of galaxy dark matter
halos. Making use of all available data, we find $\sigma_*=111\pm12$
km/s (68.3\% confidence, marginalised over the truncation parameter
$s$) for a truncated isothermal sphere model in which all galaxies
have the same mass-to-light ratio. The value of the truncation
parameter $s$ is not constrained that well, and we find
$s_*=260^{+124}_{-73}~h^{-1}$ kpc (68.3\% confidence, marginalised
over $\sigma_*$), with a 99.7\% confidence lower limit of $80~h^{-1}$
kpc. Interestingly, our results provide a 95\% confidence upper limit
of $556~h^{-1}$ kpc. The galaxy-galaxy lensing analysis allows us to
estimate the average mass-to-light ratio of the field, which can be
used to estimate $\Omega_m$. The current result, however, depends
strongly on the assumed scaling relation for $s$.

\vskip 0.05in
\noindent
{\it Subject headings:} cosmology: observations $-$ dark matter $-$ 
gravitational lensing
\end{abstract}

\vskip 0.2in
\section{Introduction}

\footnotetext[1]{Based on observations made with the William Herschel
Telescope operated on the island of La Palma by the Isaac Newton Group
in the Spanish Observatorio del Roque de los Muchachos of the
Instituto de Astrofisica de Canarias.}

The (small) differential deflection of light rays by intervening
structures allows us to study the projected mass distribution of the
deflectors, without having to rely on assumptions about the state or
nature of the deflecting matter. The first attempt to detect this
effect, called weak gravitational lensing, was made by Tyson et
al. (1984), who tried to measure the signal induced by an ensemble of
galaxies. This area of astronomy blossomed with the successful
measurements of the signal induced by rich clusters of galaxies at
intermediate redshifts (e.g., Tyson, Wenk, \& Valdes 1990; Bonnet,
Mellier, \& Fort 1994; Fahlman et al. 1994; Squires et al. 1996;
Luppino \& Kaiser 1997; Hoekstra et al. 1998; for an extensive review
see Mellier 1999). 

These studies of rich clusters were an important first step in
demonstrating the feasibility of weak lensing analyses, but nowadays
more and more studies concentrate on blank fields. For example, galaxy
groups have masses intermediate between clusters of galaxies and
galaxies. Hoekstra et al. (2001) measured the ensemble averaged weak
lensing signal from a sample of 50 groups identified by Carlberg et
al. (2001) in the Canadian Network for Observational Cosmology Field
Galaxy Redshift Survey (CNOC2).

Other applications of wide field lensing are the measurement of the
lensing signal caused by large scale structure (Bacon et
al. 2000,2002; Hoekstra et al. 2002a, 2002b; Kaiser, Wilson, \&
Luppino 2000; Refregier et al. 2002; van Waerbeke et al. 2000, 2001,
2002; Wittman et al. 2000), and the study of galaxy biasing (Hoekstra,
Yee \& Gladders 2001b; Hoekstra et al. 2002c). Another important
application is the study of the dark matter halos of field galaxies
(e.g., Brainerd, Blandford, \& Smail 1996; Griffiths et al. 1996;
Dell'Antonio \& Tyson 1996; Hudson et al. 1998; Fischer et al. 2000;
Wilson, Kaiser, \& Luppino 2001; McKay et al. 2001; Smith et al. 2001).

Rotation curves of spiral galaxies have provided important evidence
for the existence of dark matter halos (e.g., van Albada \& Sancisi
1986). Also strong lensing studies of multiple imaged systems require
massive halos to explain the oberved image separations. However, both
methods provide mainly constraints on the halo properties at
relatively small radii. The weak lensing signal can be measured out to
large projected distances, and in principle it can be a powerful probe
of the potential at large radii, constraining the extent of the dark
matter halos (e.g., Brainerd et al. 1996, Hudson et al. 1998; Fischer
et al. 2000). Only satellite galaxies (e.g., Zaritsky \& White 1994)
provide another way to probe the outskirts of isolated galaxy halos.

The lensing signal induced by an individual galaxy is too low to be
detected, and one has to study the ensemble averaged signal around a
large number of lenses. Redshifts for the individual galaxies are
useful, because they allow a proper scaling of the lensing signal
around the galaxies, and they are necessary for studies of the
evolution of the mass-to-light ratio of field galaxies from lensing.
Hudson et al. (1998) were the first to make use of (photometric)
redshifts in their galaxy-galaxy lensing analysis of the northern
Hubble Deep Field.  Unfortunately, the small area covered by the HDF
limited the accuracy of their results.

The analysis of commissioning data of the Sloan Digital Sky Survey
(SDSS) by Fischer et al. (2000) was a major step forward. Fischer et
al.  (2000) detected a very significant lensing signal, demonstrating
the importance of the survey for the study of galaxy halos. More
recently, McKay et al. (2001) used the available redshift information
from the SDSS to study the galaxy-galaxy lensing signal as a function
of galaxy properties.

We obtained deep $R$-band imaging data for two fields that have been
studied as part of the CNOC2 Field Galaxy Redshift Survey (e.g., Yee
et al. 2000). Earlier results on groups of galaxies, based on these
data, were presented by Hoekstra et al. (2001a). In this paper, we use
the data to study the galaxy-galaxy lensing signal of three
overlapping samples of galaxies (one with, and two without redshift
information).

The structure of the paper is as follows. In Section~2 we present the
observations and data reduction. In this section we also describe in
detail the object analysis and the corrections for the various
observational distortions.  In Section~3 we discuss the redshift
distribution of the sources we use in this study. We investigate the
lensing by large scale structure in Section~4. The analysis of the
galaxy-galaxy lensing signal is presented in Section~5. In Section~6
we present our estimates of the field mass-to-light ratio for
different halo models. Throughout the paper we take $H_0=100 h$
km/s/Mpc, $\Omega_m=0.2$, and $\Omega_\Lambda=0$, although the results
do not depend critically on the adopted cosmology.

\section{Data}

The Canadian Network for Observational Cosmology Field Galaxy Redshift
Survey (CNOC2) targeted four widely separated patches on the sky to
study the field population of galaxies in the universe. Redshifts of
$\sim 6200$ galaxies with a nominal limit of $R_c=21.5$ were 
measured, resulting in a large sample of galaxies at intermediate
redshifts $(z=0.12-0.55)$. A detailed description of the survey,
and the corresponding data reduction is given in Yee et al. (2000). 
In this paper we study the dark matter properties of the galaxies
targeted by CNOC2, and to this end we make extensive use of the 
redshifts and multi-colour photometry obtained by the CNOC2 survey.

We observed the central parts of the two CNOC2 patches 1447+09 and
2148-05 (Yee et al. 2002, in preparation) using the 4.2m William
Herschel Telescope (WHT) at La Palma. The images were taken using the
prime focus camera, equipped with a thinned $2048\times4096$ pixels
EEV10 chip, with a pixel scale of {0}\farcs{237} pixel$^{-1}$. The
resulting field of view of the camera is approximately {8}\farcm{1} by
{16}\farcm{2}. 

The patches observed in the CNOC2 survey are much larger than the
field of view of the WHT prime focus camera, and we observed a mosaic
of 6 pointings.  Table~\ref{fields} lists the central positions of the
observed fields as well as the dates of the observations. The typical
integration time per pointing is one hour in $R$ (see
Table~\ref{info}).

\begin{table}
\begin{center}
\begin{tabular}{lccc}
\hline
\hline
field   & RA    & DEC & date \\
1447    & $14^h 47^m 04.3^s$  & $09^\circ 12' 39''$  & May 19 - 21 1998 \\
2148    & $21^h 48^m 34^s$    & $-05^\circ 56' 00''$ & Aug 30 - Sep 2 1997 \\
\hline
\hline
\end{tabular}
\begin{small}
\caption{Positions of the centres of the fields used for the weak lensing
analysis. The last column gives the dates of the observations.
\label{fields}}
\end{small}
\end{center}
\end{table}

\subsection{Data reduction}

The images were flatfielded, using a master flatfield constructed from
the science exposures. The images were calibrated using observations
of standard stars from Landolt (1992). 

The data for each pointing typically consists of 3 exposures of 1200s,
which were taken with small offsets. Table~\ref{info} lists the total
integration time, seeing, and number of detected objects for both
observed fields. Note that because of adopted weighting scheme for the
weak lensing analysis, the ``effective'' number of galaxies is
approximately 30\% of the total number.  The data for the 1447 field
have a median seeing of $\sim 0\farcs7$, whereas the data for the 2148
field are somewhat worse, with a seeing of $0\farcs85$.

The exposures had to be remapped before the images were combined into
the final image, because of focal plane field distortions. We selected
and measured the positions of stars in each of the exposures and used
these as input for the IRAF tasks {\tt geomap} and {\tt geotran}. To
obtain the final images that were used for the object analysis, the
remapped images were simply averaged to ensure that neither cosmic ray
rejection or medianing changed the shape of the PSF or the galaxies in
a non-linear way.

Although a remapping of the images was necessary, the camera induced
distortion is small. The WHT prime focus observer's manual lists the
coefficents to estimate the telescope distortion, which is purely
radial. We also calculated the distortion from a comparison of
positions of bright stars that coincide with stars from the USNO
catalog. The relatively large uncertainty $(\sim 0\farcs3)$ in the
astrometry of the USNO catalog limits the accuracy of this approach.
The results, however, agree with the distortion derived from the
parameters listed in the WHT observer's manual, and we use the latter
to calculate the camera distortion used in our analysis.  We find that
the induced shear is small: at most 0.66\% in the corners of the
field. The corrected weak lensing distortion field is obtained by
subtracting the camera distortion from the observed distortions (which
are corrected for PSF anisotropy), as described in Hoekstra et
al. (1998).

\begin{table}
\begin{center}
\begin{tabular}{lccccc}
\hline
\hline
field	& $t_{\rm exp}$ & seeing & \# objects & $\bar n$ & $m_{\rm lim}$ \\
	& [s] & [''] &  & [arcmin$^{-2}$] & (90\%)\\
(1) 	&(2)	&(3)	&(4)	&(5)	&(6) \\
\hline
1447-1	& 6000	& 0.75	& 5738	& 46	& 24.3	\\
1447-2	& 3600	& 0.70	& 5545	& 44	& 24.5	\\
1447-3	& 3600	& 0.65	& 6419	& 51	& 24.8	\\
1447-4	& 3600	& 0.70	& 5570	& 45	& 24.6	\\
1447-5	& 4800	& 0.70	& 6087	& 49	& 24.5	\\
1447-6	& 3600	& 0.65	& 6826	& 55	& 24.8	\\
\hline			
2148-1	& 3600	& 1.0	& 4712	& 38	& 24.2	\\
2148-2	& 3600	& 0.80	& 4976	& 40	& 24.3	\\
2148-3	& 3600	& 0.75	& 5540	& 44	& 24.2	\\
2148-4	& 4800	& 0.85	& 4804	& 38	& 24.2	\\
2148-5	& 3600	& 0.90	& 4486	& 36	& 24.1	\\
2148-6	& 3600	& 0.85	& 5110	& 41	& 24.2	\\
\hline
\hline
\end{tabular}
\begin{small}
\caption{Summary of the deep WHT imaging. Column~(1): identification of the 
pointing; (2) total integration time; (3) median seeing; (4) number of 
galaxies; (5) number density of galaxies; (6) 90\% completeness limit in $R$.
\label{info}}
\end{small}
\end{center}
\end{table}

\subsection{Object detection and analysis}

Our analysis techique is based on that developed by Kaiser, Squires, \& 
Broadhurst (1995) and Luppino \& Kaiser (1997), with a number of modifications
which are described in Hoekstra et al. (1998). We analyse each pointing 
separately as the seeing and PSF anisotropy vary between each exposure. 
After the catalogs have been corrected for the various observational effects, 
they are combined into a master catalog which covers the complete field that 
is observed.

The first step in the analysis is to detect the faint galaxy images,
for which we used the hierarchical peak finding algorithm from Kaiser
et al.  (1995). The peak finder also provides an estimate for the
Gaussian scale length $r_g$ of the object. We select objects which
were detected with a significance $\nu>5\sigma$ over the local sky in
the combined images.  This galaxies are used for the weak lensing
analysis. We also run the peak finder on the images of the single
exposures. We do a coincidence test on these catalogs and classify
extremely small, but very significant objects as cosmic rays, which
are removed from the catalog that is used for the object analysis.

Some faint cosmic rays may hit galaxies, and consequently might not be
recognized as cosmic rays. Based on the number of cosmic ray hits, and
the area covered by galaxies we find that less than $0.2\%$ of the
galaxies might be affected. Also, cosmic rays only introduce
additional noise in the shape measurement, but do not bias the
result. Consequently we conclude that remaining cosmic rays have a
negligble effect on our results.

For the weak lensing analyis we select objects which are detected in
at least two of the shorter exposures. The objects which are detected
in only one of the shorter exposures are small faint objects, which
are not useful as their shape parameters are noisy. The resulting
catalogs are inspected visually in order to remove spurious
detections, such as spikes from saturated stars, HII regions in
resolved galaxies, etc.

For all detected objects we measure the apparent magnitude in an
aperture with a radius of 3 times the Gaussian scale length of the
object, the half light radius, and the shape parameters (polarization
and polarizabilities). We also estimate the error on the polarization
following Hoekstra, Franx, \& Kuijken (2000).

\subsection{PSF correction}

To measure the small, lensing induced, distortions in the images of
the faint galaxies it is important to correct the shapes for
observational effects, such as PSF anisotropy and seeing: PSF
anistropy can mimic a lensing signal, and the correction for seeing is
required to relate the measured shapes to the real lensing signal.

\begin{figure*}
\begin{center}
\leavevmode
\hbox{%
\epsfxsize=\hsize
\epsffile[23 350 590 715]{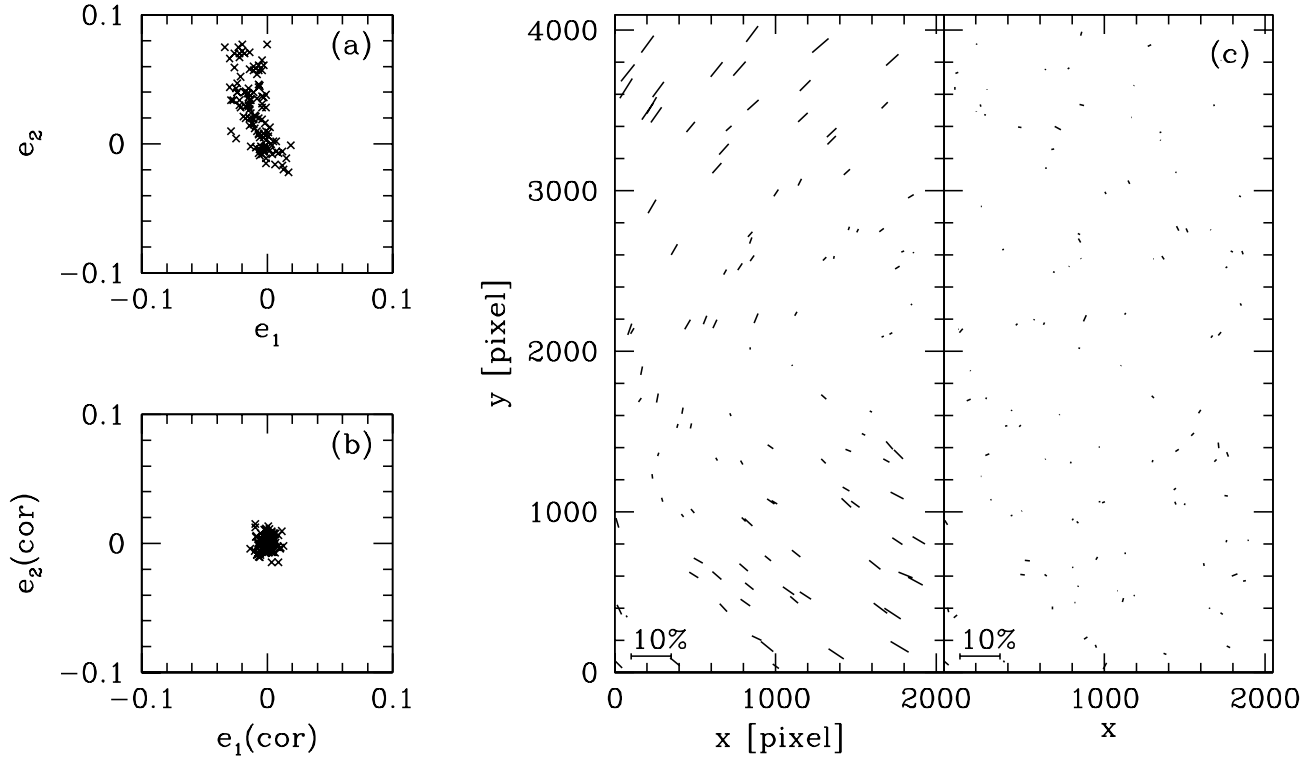}}
\begin{small}
\caption{Shape of the PSF in one of the pointings of the 1447 field; (a)
the polarizations of the stars before correction; (b) residual polarization
after correcting for the PSF anisotropy using a second order polynomial
fit to the data. (c) PSF anisotropy as a function of position on the chip.
The left panel shows the PSF anisotropy before correction. To show the higher 
order components the average polarization has been subtracted. The direction 
of the sticks indicates the major axis of the PSF, whereas the length 
corresponds to the size of the anisotropy. The right panel shows the 
residuals in the stellar polarizations after correction.
\label{psf}}
\end{small}
\end{center}
\end{figure*}

To do so, we follow the procedure outlined in Hoekstra et al. (1998)
(which is based on Kaiser et al. 1995; Luppino \& Kaiser 1997). We
select a sample of moderately bright stars from our
observations. These are used to characterize the PSF anisotropy and
seeing. Figure~\ref{psf} shows a typical result for one of the
pointings. We fit a second order polynomial to the observed shape
parameters of the stars, and this model is used to correct for the PSF
anisotropy.  Figure~\ref{psf}b also shows the residual polarization of
the stars after the correction. The residuals are small indicating
that we can reliably correct for the PSF anisotropy.

\begin{figure}
\begin{center}
\leavevmode
\hbox{%
\epsfxsize=\hsize
\epsffile[0 150 560 690]{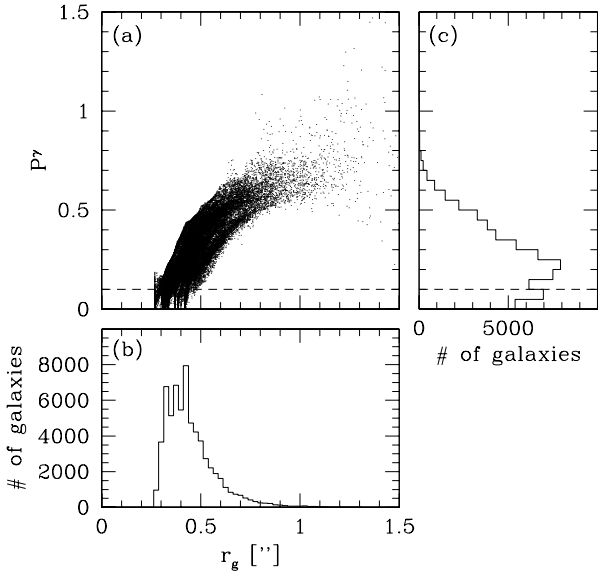}}
\begin{small}
\caption{(a) $P^\gamma$ as a function of $r_g$ for all galaxies. Due to
seeing variations the curves for the various fields blend; (b) number
of galaxies with a given $r_g$; (c) number of galaxies with a given
value of $P^\gamma$. For the weak lensing analysis we restrict the sample
to galaxies with $P^\gamma\ge 0.1$.
\label{pgam}}
\end{small}
\end{center}
\end{figure}

The next step is to correct the shapes for the circularization by the
PSF.  The stars that were used to study the PSF anisotropy are also
used to compute the `pre-seeing' shear polarizability $P^\gamma$
(Luppino \& Kaiser 1997; Hoekstra et al. 1998). The measurement of
$P^\gamma$ is very noisy, and we combine the estimates of many
galaxies to reduce the noise. We also found that the size of the PSF,
and thus the correction, depends on the position on the chip. The
variation, however, is small: about 10\% maximum. To account for this,
we bin the raw polarizabilities not only in bins of $r_g$, but also as
a function of position. For a given $r_g$ we fit a second order
polynomial to the median $P^\gamma$'s as a function of position, and
use the results to compute the $P^\gamma$ for each
galaxy. Figure~\ref{pgam}a shows $P^\gamma$ as a function of $r_g$.

Objects with small values for $P^\gamma$ require large corrections,
thus increasing the noise. The weighting scheme suggested in Hoekstra
et al. (2000) gives already less weight to these objects, but in addition
we exclude objects with $P^\gamma\le0.1$

\section{Redshifts of lenses and sources}

For a given mass distribution, the amplitude of the weak lensing
signal is proportional to the dimensionless mass surface density 

\begin{equation}
\kappa(\vec x)=\frac{\Sigma(\vec x)}{\Sigma_{\rm crit}},
\end{equation}

\noindent where the critical surface density is defined as
\begin{equation}
\Sigma_{\rm crit}=\frac{c^2}{4\pi G}\frac{D_{\rm s}}{D_{\rm l} D_{\rm ls}}
=\frac{c^2}{4\pi G D_{\rm l} \beta}.
\end{equation}

\noindent Here $D_{\rm s}$, $D_{\rm l}$, and $D_{\rm ls}$ correspond
to the angular diameter distances between the observer and the source,
observer and the lens, and the lens and the source. The lensing
signal depends on both the redshifts of the lenses and the sources,
and the dependence on the source redshift is characterized by $\beta$, which
is defined as

\begin{equation}
\beta=\max[0,D_{\rm ls}/D_{\rm s}].
\end{equation}

If the redshift of the lens approaches that of the source, the
amplitude of the lensing signal, which is proportional to $\beta$,
decreases. 

Most of the lensing signal comes from galaxies that are generally too
faint to be included in redshift surveys. To relate the lensing signal
to a physical mass it is necessary to know the redshift distribution
of the faint background galaxies. To this end, we use photometric
redshift distributions from the Hubble Deep Fields North and South
(Fern{\'a}ndez-Soto, Lanzetta, \& Yahil 1999).
Hoekstra et al. (2000) used these results for their analysis of the
distant cluster MS~1054-03 $(z=0.83)$, and concluded that these
redshift distributions provide a good approximation of the true
distribution. We use the colours of the galaxies in the HDFs to derive
their $R$ band magnitude. For each lens-source pair we compute the
corresponding value of $\beta$.

The lenses considered in our analysis are at much lower redshifts
than MS~1054-03, and consequently the uncertainty in the value of
$\beta$ is small. Based on field-to-field variation in the
redshift distribution and the uncertainty due to the finite number
of galaxies in the Hubble Deep Fields, we estimate that the uncertainty
in $\beta$ is 2\%.

\section{Lensing by large scale structure}

One of the selection criteria for the CNOC2 fields (which are
described in Yee et al. 2000) is that no known nearby rich cluster
should be in the field. However, the observed fields might contain
distant massive clusters. In principle, such clusters can be found in
a weak lensing analysis, provided they are massive enough (e.g.,
Wittman et al. 2001), even if they are ``dark'' (see Erben et
al. 2000). We note that projection effects can actually introduce
spurious detections (e.g., Hoekstra 2001; White, van Waerbeke \&
Mackey 2002).

To investigate possible structures in the CNOC2 fields, we
reconstructed maps of the dimensionless surface density, using the
original Kaiser \& Squires (1993) algorithm.  The resulting mass maps
are consistent with noise maps. We find a few $3\sigma$ peaks, but no
obvious counterparts are seen in the number counts of bright
galaxies. The mass maps have been smoothed with a Gaussian with a FWHM
of 1 arcminute, resulting in approximately 500 independent points in
each map. Therefore one expects about three $3\sigma$ peaks.

It is useful to estimate the mass detection limit for our data.  We
assume that the cluster mass profile is well described by a singular
isothermal sphere (SIS):

\begin{equation}
\kappa=\frac{r_E}{2r},
\end{equation}

\noindent where $r_E$ is the Einstein radius. The observed scatter in
the shapes of the sources results in a typical uncertainty in the
Einstein radius $r_E$ of $\sim 2''$.  We fit SIS models out to 5
arcminutes, but note that the uncertainty in the determination of the
Einstein radius does not depend significantly on this particular
range. Hence, we should be able to detect a cluster with $r_E>6''$ (at
the $3\sigma$ level). The velocity dispersion of the SIS model is
given in km/s by

\begin{equation}
\sigma=186.3\sqrt{\frac{r_E}{\beta}}~{\rm km/s},
\end{equation}

\noindent when $r_E$ is given in arcseconds.  Given the redshift
distribution of our sources a cluster with a velocity dispersion of
630 km/s at $z=0.2$ would be detectable, whereas a cluster at $z=0.5$
would have to have a velocity dispersion larger than 900 km/s. Such
rich clusters would have been detected in the CNOC2 redshift survey. A
detectable cluster at $z>0.5$ would have to be even more massive. Such
massive systems are very rare, and are unlikely to be in the observed
fields. Based on the mass reconstructions we find no massive clusters
in the CNOC2 fields.

Although we do not detect significant structures in the CNOC2 field,
any large scale structure along the line of sight will introduce an
excess alignment in the shapes of the faint galaxies (compared to a
random field). The measurement of this signal provides a direct
measurement of the statistical properties of the large scale mass
distribution (e.g., Blandford et al 1991; Kaiser 1992; Bernardeau, van
Waerbeke, \& Mellier 1997; Schneider et al. 1998).

\begin{table}
\begin{center}
\begin{tabular}{lccc}
\hline
\hline
field   & $\langle g_1 \rangle$	& $\langle g_2 \rangle$ & $\sigma_g$ \\
1447    & -0.00334   		&  0.00144		&   0.00226 \\
2148    &  0.00323   		& -0.00140     		&   0.00254 \\
\hline
\hline
\end{tabular}
\begin{small}
\caption{Average distortion in the observed fields. The errors in
the measurements are given in the last column. The values are small,
which suggest that the correction for the PSF anisotropy has worked
well.
\label{avshear}}
\end{small}
\end{center}
\end{table}

\begin{figure}
\begin{center}
\leavevmode
\hbox{%
\epsfxsize=8cm
\epsffile{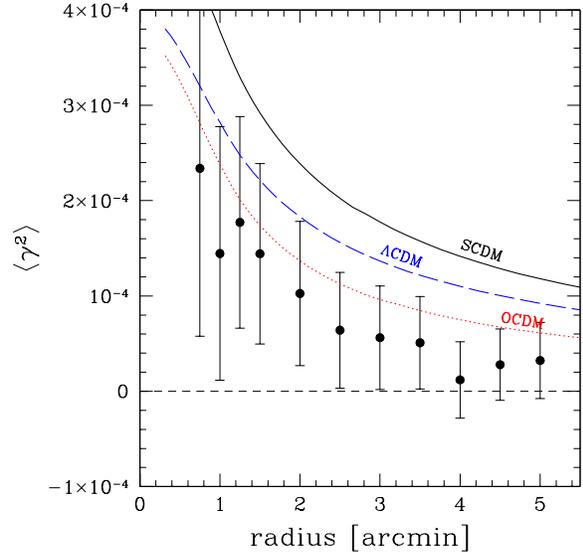}}
\begin{small}
\caption{The observed variance of $\langle\gamma^2\rangle$ as a
function of the radius of the aperture in which the shear is
averaged. Note that the points are correlated.  The error bars do not
include the additional uncertainty caused by cosmic variance.  For
reference we have displayed the expected variances for three different
cosmologies: SCDM $(\Omega_m=1.0; \Omega_\Lambda=0; \Gamma=0.5;
\sigma_8=0.5)$, OCDM $(\Omega_m=0.3; \Omega_\Lambda=0; \Gamma=0.21;
\sigma_8=0.85)$, and $\Lambda$CDM $(\Omega_m=0.3; \Omega_\Lambda=0.7;
\Gamma=0.21; \sigma_8=0.9)$
\label{cosmic}}
\end{small}
\end{center}
\end{figure}

A widely used method is to look for the excess variance
in an aperture: the top-hat variance.  This measurement as a function
of scale can be compared to predictions from cosmological models. The
measurement is difficult, and residual systematics (such as an
imperfect correction for PSF anisotropy) increase the observed
variance.

Since the first detections (Bacon et al. 2000; Kaiser, Wilson, \&
Luppino 2000; Maoli et al. 2001; van Waerbeke et al. 2000; Wittman et
al. 2000) tremendous progress has been made. The most recent results
are based on large data sets, and yield consistent results, despite
the small statistical errors (e.g., Bacon et al. 2002; Hoekstra et
al. 2002a, 2002b; Refregier et al.  2002; van Waerbeke et al. 2001,
2002). The area covered by the two CNOC2 fields is smaller than the
area covered by these studies, but it is still interesting to examine
our results and compare them to the predictions: although the analysis
will not provide tight constraints on the cosmology, it is a useful
test whether residual systematics are present in the data. 

For the analysis, we use galaxies with apparent magnitudes $21<R<26$
as our sample of sources. The measurements of the average distortion
of the two fields are presented in Table~\ref{avshear}. To measure the
excess top-hat variance as a function of aperture radius, we tile the
observed fields with apertures of a given scale, compute the variance,
and subtract the contribution from the intrinsic shapes of the
galaxies (e.g., Bacon et al. 2000; Kaiser et al. 2000; Hoekstra et
al. 2002a; van Waerbeke et al. 2000, 2001). 

The results are presented in Figure~\ref{cosmic}. The error only
includes the statistical uncertainty caused by the intrinsic shapes of
the sources, and does not include the contribution from cosmic
variance. Also note that the points are correlated. The 
measured variances are small, and consistent with current, more
accurate measurements.

The top-hat variance is very sensitive to residual PSF anisotropy, and
can be used to test the accuracy of the correction. For instance, if
we do not correct the shapes of the galaxies for PSF anisotropy, we
measure an excess variance $\langle\gamma^2\rangle= (8.4\pm0.6)\times
10^{-3}$ at a scale of 1 arcminute, 60 times larger than the signal
presented in Figure~\ref{cosmic}. If we subtract 90\%
(under-correction) or 110\% (over-correction) of the PSF anisotropy,
the observed variance increases by a factor $\sim 2$. Because any
residual PSF anisotropy will increase the signal, this test indicates
that the adopted correction yields the minimum (or close to) variance.

\section{Galaxy-galaxy lensing}

We use two different methods to study the lensing signal caused by the
field galaxies. First we measure the ensemble averaged tangential
distortion around the lens galaxies (galaxy-mass correlation
function), which provides a robust estimate of the lensing signal. 
We also perform a maximum likelihood analysis, which is described in
Section~5.3.

It is important to note that the measurement of the galaxy-galaxy
lensing signal is much less sensitive to a small residual systematics:
we measure the lensing signal that is perpendicular to the line
connecting many lens-source pairs. These connecting lines are randomly
oriented with respect to the PSF anisotropy, and hence suppress any
residual systematics. Given the results obtained for the top-hat
smoothed variance (see Section~4), we expect that the effect of
residual systematics is negligible.

In Table~\ref{sigmatab} we present the three subsamples of lenses that
we study. The `faint' and the `bright' sample are magnitude limited,
and we use statistical redshift distributions for the lenses to infer
the average halo properties. The `CNOC2' sample has the same magnitude
limits as the `bright' sample, but consists only of galaxies with
spectroscopic redshifts. It includes approximately half of the
galaxies of the `bright' sample. The `faint' sample is comparable to
the sample of lenses studied by Brainerd et al. (1996) who used
$20<r_{\rm lens}<23$, and $23<r_{\rm source}<24$. Brainerd et
al. (1996) did not use fainter sources, because they did not correct
for observational distortions.

\begin{figure*}
\begin{center}
\leavevmode
\hbox{%
\epsfxsize=8cm
\epsffile{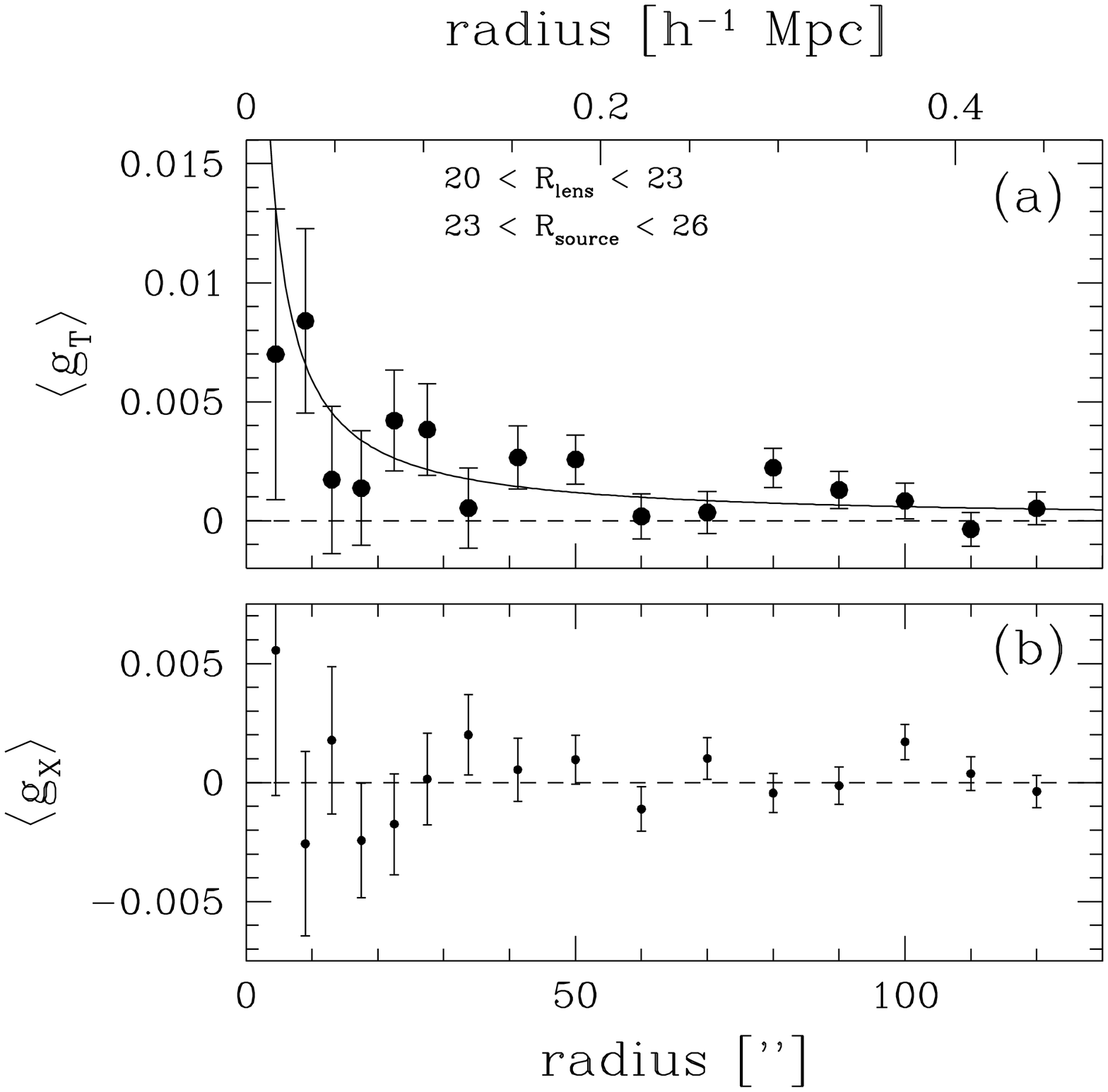}
\epsfxsize=8cm
\epsffile{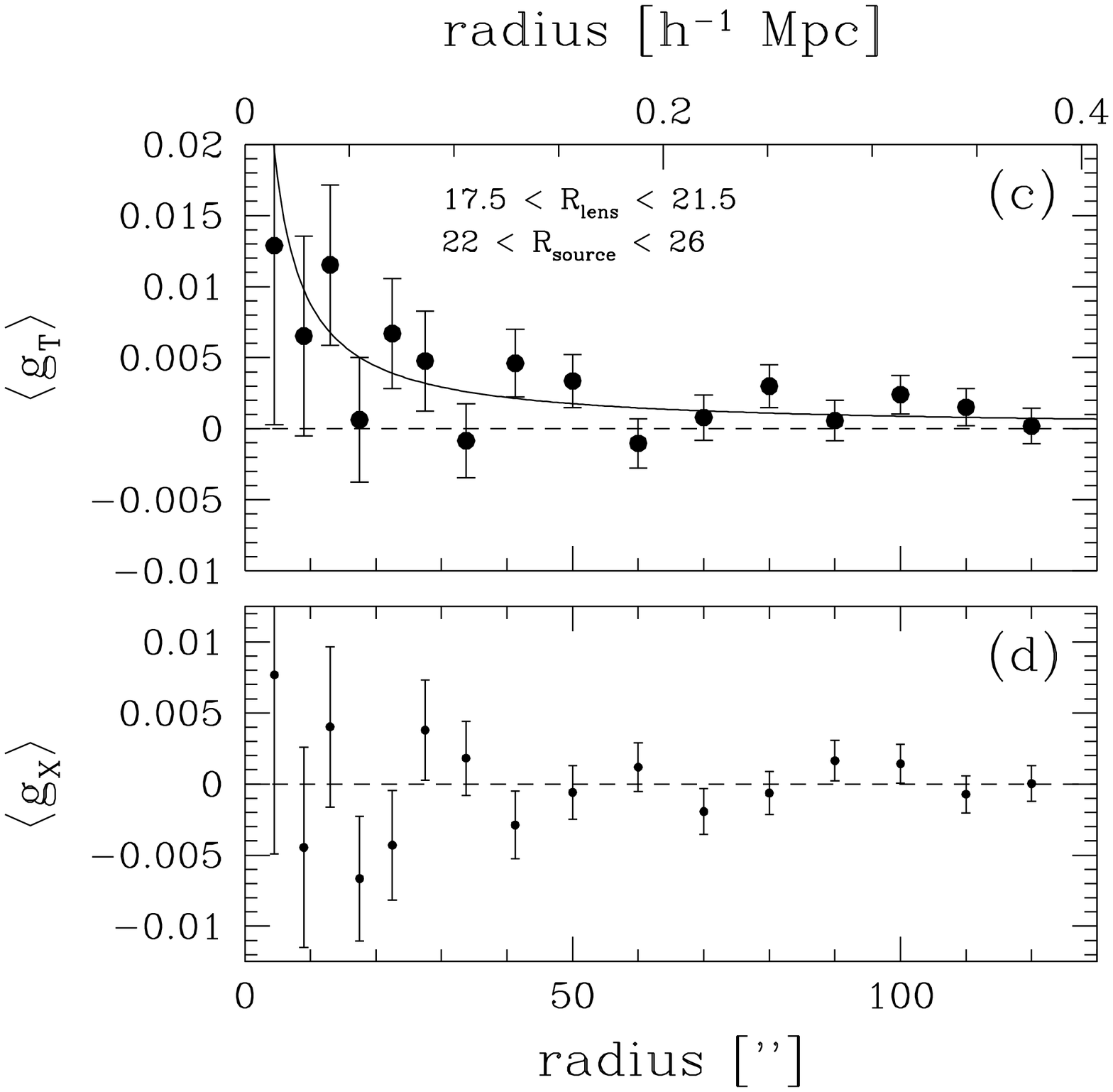}}
\begin{small}
\caption{(a) The ensemble averaged tangential distortion for the
`faint' sample. The solid line corresponds to the best SIS model fit,
which has $r_E=0\farcs{118}\pm0\farcs{025}$. Panel~c shows the signal
around the `bright' sample, and we find
$r_E=0\farcs{176}\pm0\farcs{045}$.  Panels~b and~d show the average
signal when the sources are rotated by $\pi/4$. No signal should be
present if the signal in panels~a and~c is caused by lensing.
\label{galgalnoz}}
\end{small}
\end{center}
\end{figure*}

\begin{table*}
\begin{center}
\begin{tabular}{ccccccccccc}
\hline
\hline
 (1)	& (2)	& (3)	& (4) & (5) & (6) & (7) & (8) & (9) & (10) & (11) \\
sample	& lens	& source & \# lens & $z_{\rm lens}$ & $\langle\beta\rangle$ & $\langle r_E\rangle$ & $\langle r_E^*\rangle$ & $\sigma_*$ & $V_c^*$ & $\sigma_*$ \\
	&	& 	 &	   &	   		   &      	           & [''] 		  & ['']  & [km/s] & [km/s] & [km/s]  \\
\hline
faint	& $20<R<23$     & $23<R<26$ & 8715 & 0.46 & $0.35\pm 0.01$ & $0.118\pm0.026$ & $0.159\pm0.035$ & $126^{+13}_{-15}$ & $178^{+19}_{-21}$ & $124^{+13}_{-15}$\\
bright	& $17.5<R<21.5$	& $22<R<26$ & 2125 & 0.34 & $0.41\pm 0.01$ & $0.176\pm0.046$ & $0.161\pm0.042$ & $117^{+15}_{-17}$ & $165^{+21}_{-23}$ & $115^{+15}_{-17}$\\
CNOC2	& $17.5<R<21.5$	& $22<R<26$ & 1125 & 0.36 & $0.40\pm 0.01$ & $0.196\pm0.047$ & $0.196\pm0.048$ & $130^{+15}_{-17}$ & $184^{+22}_{-25}$ & $129^{+15}_{-17}$\\
\hline
\hline
\end{tabular}
\begin{small}
\caption{Properties and results for the different samples of lens galaxies.
(2) the range in apparent magnitude of the lens galaxies; (3) range in
apparent magnitude for the sources; (4) number of lens galaxies; (5)
median redshift of the lenses; (6) average value of $\beta$ based on
the redshift distributions of the lenses and the sources. (7) best
fit Einstein radius; (8) estimate for the Einstein radius of an
\lstar\ galaxy; (9) best estimate for the velocity dispersion
of an \lstar\ galaxy under the assumption that the luminosity
evolves $\propto(1+z)$; (10) corresponding circular velocity;
(11) the velocity dispersion of an \lstar\ galaxy for no
luminosity evolution.
\label{sigmatab}}
\end{small}
\end{center}
\end{table*}  

\subsection{Lenses selected irrespective of redshift information}

The ensemble averaged tangential distortion as a function of radius
from the lens is a well established way to present the galaxy-galaxy
lensing signal (e.g., Brainerd et al. 1996; Hudson et al. 1998;
Fischer et al. 2000; McKay et al. 2001). The results for the `faint'
and the `bright' sample are shown in Figure~\ref{galgalnoz}.  For both
samples a significant lensing signal is detected. If the measured
signal is caused by gravitational lensing, no signal should be
observed when the sources are rotated by $\pi/4$.  The results of this
test are shown in figure~\ref{galgalnoz}b and~d, and no signal is seen
indeed.

If the lenses are distributed randomly on the sky, the tangential
distortion profile can be related directly to the ensemble averaged
mass profile of the lenses. In reality the lenses cluster, and the
observed signal is the convolution of the mass profile and the galaxy
correlation function: the tangential shear profile measures the
galaxy-mass cross-correlation function. As a result mass estimates
based on the tangential shear profile are biased high, because at
large distances from the lens, one measures the mass of the lens and
associated galaxies. However, the lens dominates the signal on small
scales.

We fit a SIS model to the tangential distortion profiles presented in
figure~\ref{galgalnoz}. We do not know the redshifts of the individual
lenses in these samples, and consequently we can only determine the
ensemble averaged Einstein radius $\langle r_E \rangle$.  For the
`faint' sample we find a best fit Einstein radius $\langle r_E\rangle
=0\farcs{118}\pm0\farcs{025}$, and for the `bright' sample we obtain
$\langle r_E\rangle=0\farcs{176}\pm0\farcs{045}$.  

We examined the effect of an imperfect correction for PSF anisotropy
on the determination of the Einstein radius. Even in the extreme case
that no correction is applied the derived value is changed by only
5\%. In the case of a 90\% or 110\% correction of the PSF anisotropy,
the signal is changed by $\sim 1\%$. Based on the results obtained
for the top-hat smoothed variance (see Section~4), we conclude that
the effect of an imperfect correction of the PSF anisotropy is
much smaller than 1\%, and hence is negligible.

We use the effective $\beta$ for these samples (see column 6 of
Table~\ref{sigmatab}) to derive a mass weighted average velocity
dispersion $\langle\sigma^2\rangle^{1/2}=108^{+11}_{-12}$ km/s for the
`faint' sample, and $\langle\sigma^2\rangle^{1/2}=122^{+15}_{-17}$
km/s for the `bright' sample. The corresponding circular velocity can
be calculated using $V_c=\sqrt{2}\sigma$.

The derived values of $\langle\sigma^2\rangle^{1/2}$ depend on the
selection of the sample of lens galaxies, and one cannot compare
these results to findings of other studies, given the differences
in sample selection. Instead we estimate the
velocity dispersion (or circular velocity) of an \lstar\ galaxy.  We
assume a scaling relation between the velocity
dispersion and the luminosity of the galaxy of the form

\begin{equation}
\sigma\propto L_{B}^{1/4}, {\rm~or~} V_c\propto L_B^{1/4}.
\end{equation}

We also assume that the luminosity of a lens of given mass
evolves with redshift $\propto (1+z)$ (e.g., Lin et al. 1999).
With these assumptions the average value of the Einstein radius
(in radians) is given by

\begin{equation}
\langle r_E \rangle=\frac{4\pi}{c^2} \frac{\sigma_*^2}{\sqrt{L_B^*(z=0)}}
\left\langle\beta\sqrt{\frac{L_B^{\rm obs}}{1+z}}\right\rangle,
\end{equation}

\noindent where $L_B^{\rm obs}$ is the observed intrinsic luminosity
of the galaxy. We also introduce $\langle r_E^*\rangle$

\begin{equation}
\langle r_E^*\rangle=\frac{4\pi}{c^2} \sigma_*^2 \langle\beta\rangle,
\end{equation}

\noindent which is the Einstein radius of an \lstar\ galaxy at the 
average redshift of the sample of lenses.

The redshift distributions of the sources and the `faint' lens sample
are derived from the photometric redshift distributions of the HDF
North and South ((Fern{\'a}ndez-Soto, Lanzetta, \& Yahil 1999). For
the `bright' sample we use the redshift distribution from the CNOC2
survey with the proper weighting to take into account the
incompleteness of the survey. To determine the restframe $B$
luminosities we use template spectra for a range in spectral types and
compute the corresponding passband corrections as a function of
redshift and galaxy colour (this procedure is similar to the one
described in van Dokkum \& Franx 1996).

Lin et al. (1999) have studied the field galaxy luminosity function at
intermediate redshift from the CNOC2 survey, and derived $M_{\rm
B}^*(z=0.3)=-19.18+5\log h$, which corresponds to a luminosity of
$7.3\times 10^9~h^{-2}{\rm L}_{{\rm B}\odot}$. With our assumed
luminosity evolution, this results in \lumstar. Madgwick et al. (2002)
derived $M_{\rm B}^*=-19.55+5\log h$ from the the 2dF redshift survey,
which probes lower redshifts. Hence, the CNOC2 value (Lin et al. 1999)
is rather low, in particular with our choice of luminosity evolution.
Throughout the paper we assume that the luminosity evolves
$\propto(1+z)$, but for reference we also list the results for the
no-evolution scenario, in which case ${\rm L}^*_{\rm
B}(z=0)=7.3\times10^9~h^{-2} {\rm L}_{\rm {B}\odot}~$.

Column~8 in Table~\ref{sigmatab} lists the derived Einstein radii of
an \lstar\ galaxy for the `bright' and the `faint' sample. For the
`bright' sample the observed $\langle r_E\rangle$ is actually very
close to $r_E^*$. For the `faint' sample we find that the observed
$\langle r_E\rangle=0.74\langle r_E^*\rangle$. Columns~9 and 10 list
the velocity dispersion and circular velocity of an \lstar\ galaxy
under the assumption that the luminosity evolves
$\propto(1+z)$. Column~11 shows the resulting velocity dispersion
without luminosity evolution. Coincidentally, the values for
$\sigma_*$ for the evolving and non-evolving case are very similar.
Furthermore, the results for the faint and bright sample agree well
with one another.

\subsection{Lenses with redshifts from CNOC2}

\begin{figure}
\begin{center}
\leavevmode
\hbox{%
\epsfxsize=8cm
\epsffile{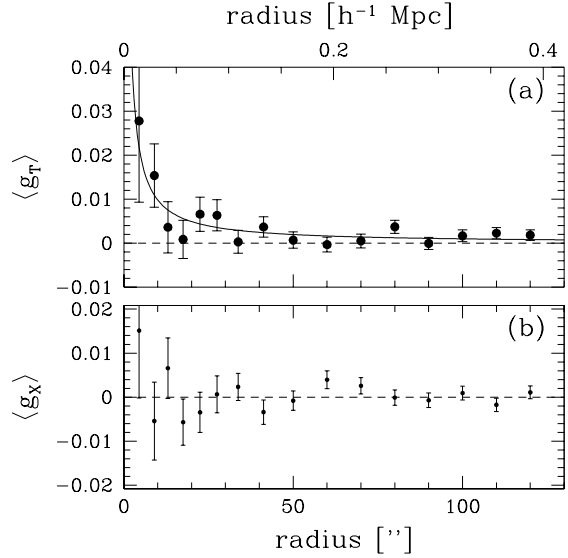}}
\begin{small}
\caption{(a) The ensemble averaged tangential distortion around
galaxies with $17.5<R<21.5$ and measured spectroscopic redshifts,
scaled to an $L_B^*$ galaxy at a redshift $z=0.36$. The solid line
corresponds to the best SIS model fit, which has
$r_E=0\farcs{196}\pm0\farcs{047}$. (b) The average signal when the
sources are rotated by $\pi/4$. No signal should be present if the
signal in panel~a is caused by lensing.
\label{galgalz}}
\end{small}
\end{center}
\end{figure}

Redshifts for the lens galaxies are useful because they allow a proper
scaling of the signals around the lens galaxies.  Intrinsically faint
(and therefore low mass) galaxies, or galaxies with redshifts
comparable to the source galaxies are given lower weights.  We scale
the observed distortion of each galaxy, as well as its error, such that
it corresponds to that of an \lstar\ galaxy at the median redshift of
the `CNOC2' sample

\begin{equation}
g_T^{\rm scale}=\left(\frac{5.6\times 10^9(1+z)}{L_B}\right)^{1/2}
\left(\frac{0.4}{\beta}\right)g_T^{\rm obs},
\end{equation}

\noindent where we assumed that the luminosity scales with the fourth
power of the velocity dispersion. The value of $\beta$ is calculated
for each galaxy separately, based on its redshift, and the redshift
distribution of the sources. The sample of lenses consists of 1125
galaxies with spectroscopic redshifts, and $17.5<R_{\rm lens}<21.5$.
We note that, because of the selection of these galaxies, the sampling
in apparent magnitude is not uniform (the completeness for the fainter
galaxies is lower). Figure~\ref{galgalz}a shows the resulting average
tangential distortion as a function of radius around an \lstar\ galaxy
at $z=0.36$. A significant galaxy-galaxy lensing signal is detected.

The best fit SIS model to the scaled tangential distortion yields an
Einstein radius $r_E=0\farcs{196}\pm0\farcs{047}$. From this result we
derive an average velocity dispersion of $\sigma=130^{+15}_{-17}$ km/s
or circular velocity of $V_c=184^{+22}_{-25}$ km/s for an \lstar\
galaxy. If we assume no luminosity evolution, we obtain
$\sigma_*=129^{+15}_{-17}$ km/s (with ${\rm L}^*_{\rm
B}(z=0)=7.3\times10^9~h^{-2} {\rm L}_{\rm {B}\odot}~$).

The number of lenses in the `CNOC2' sample is half that of the
`bright' sample, but the signal-to-noise ratio with which the lensing
signal is detected is similar. Thus redshifts for the lenses are
useful as they can be used to improve the signal-to-noise ratio of the
measurement.  A comparison of the results for the different samples
(which are listed in Table~\ref{sigmatab}) shows a good agreement.

We note that we have made assumptions about the luminosity
evolution and the scaling relations. The redshift information for the
lenses is crucial when one would like to constrain the scaling
relations or study the evolution of the mass-to-light ratio of
these field galaxies.

\subsection{Sizes of galaxy halos}

As mentioned above, the clustering of lenses will bias the mass
estimates inferred from the observed tangential distortion profile.
In this section we use a parameterised model for the mass distribution
of individual galaxies and compare the predicted distortions to the
observed distortion field. We use a truncated halo model, in order
to constrain both the mass and the extent of the dark matter halos.

This approach naturally takes into account the clustering of the
lenses in the sample. Furthermore, we make use of both components of
the distortion. It is important to note that the contribution of
associated faint galaxies is not accounted for. The effect of a
large-scale cosmological shear on the parameter estimation is
negligble because one measures the lensing signal for many
randomly oriented lens-source pairs. The effect of clusters/groups
on the results is examined in Section~5.4.

A useful model to describe a truncated halo is (Schneider
\& Rix 1997)

\begin{equation}
\Sigma(r)=\frac{\sigma^2}{2Gr}\left(1-\frac{r}{\sqrt{r^2+s^2}}\right),
\end{equation}

\noindent where $s$ is a measure of the truncation radius. The total
mass of this model is finite, and half of the mass is contained within
$r=\frac{3}{4}s$. The total mass is given by

\begin{equation}
{\rm M}_{\rm tot}=\frac{\pi \sigma^2}{G}s=
7.3\times 10^{12} \left(\frac{\sigma}{100~{\rm km/s}} \right)^2
\left(\frac{s}{1~{\rm Mpc}}\right).
\end{equation}

The scatter in the polarizations of the galaxies is approximately
constant with apparent magnitude, and it can be well approximated by a
Gaussian distribution. In that case the log-likelihood follows a
$\chi^2$ distribution with the number of degrees of freedom equal to
the number of free model parameters (e.g., Hudson et al. 1998), and
the determination of confidence intervals is straightforward. The
log-likelihood is given by the sum over the two components of the
polarization $e_i$ of all the source galaxies

\begin{equation}
\log{\cal L}=-\sum_{i,j} \left(\frac{e_{i,j} - g_{i,j}(\sigma_*,s_*) 
P^\gamma_j}{\sigma_{e_j}}\right)^2,
\end{equation}

\noindent where $g_{i,j}$ are the model distortions, $P^\gamma_j$
is the shear polarizability, $e_{i,j}$ are the image polarizations
for the $j$th galaxy, and $\sigma_{e_j}$ is the uncertainty in the
shape measurement of the $j$th galaxy.

In our maximum likelihood analysis we ignore the contribution from
lenses outside the field of view (e.g., Hudson et al. 1998). For small
fields of view this tends to lower the resulting $\sigma_*$ and $s_*$
somewhat. The area covered by our observations is much larger than the
HDF North studied by Hudson et al. (1998), and we find that the effect
on our estimates of $\sigma_*$ and $s_*$ is negligible.

To scale the signals of the various lenses we use again $\sigma\propto
L_B^{1/4}$, which is based on both dynamical and observational
considerations. The situation is different for the truncation
parameter $s$, because there are no observational constraints. Here we
will explore several cases.

If all halos have the same value for $s$ the total
mass-to-light ratio scales as $({\rm M/L})_{\rm tot}\propto L^{-1/2}$
(where we assume that $L\propto \sigma^4$). Thus the mass-to-light
ratios of more luminous galaxies are lower.  Another option is to take
$({\rm M/L})_{\rm tot}={\rm constant}$ for all galaxies. This choice
is equivalent to taking $s\propto \sigma^2$ (e.g., Brainerd et
al. 1996; Hudson et al. 1998). The last relation we examine is
$s\propto \sigma^4$, which gives $({\rm M/L})_{\rm tot}\propto
L^{1/2}$.

\begin{figure*}
\begin{center}
\leavevmode 
\hbox{%
\epsfxsize=8cm \epsffile{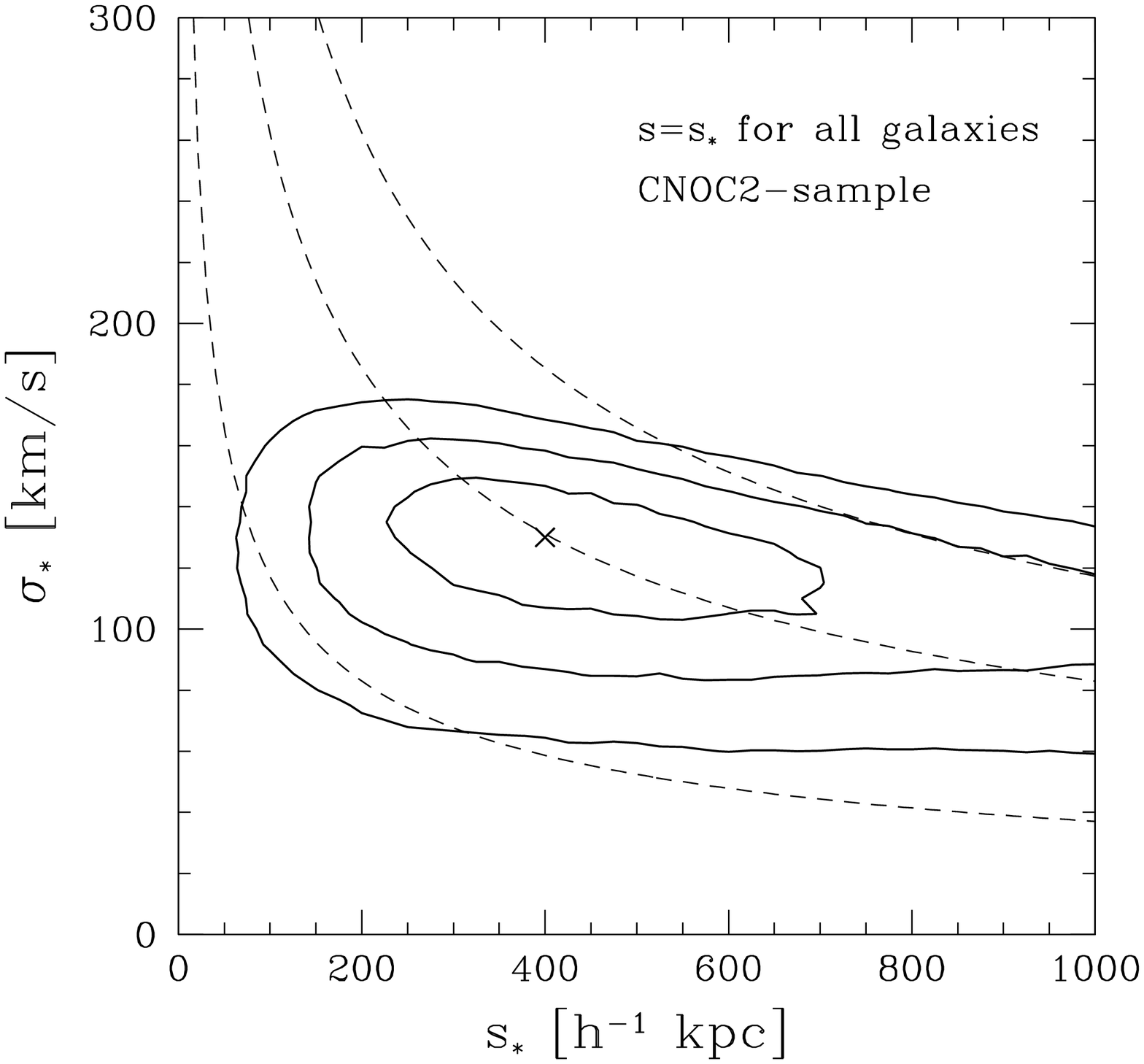}
\epsfxsize=8cm \epsffile{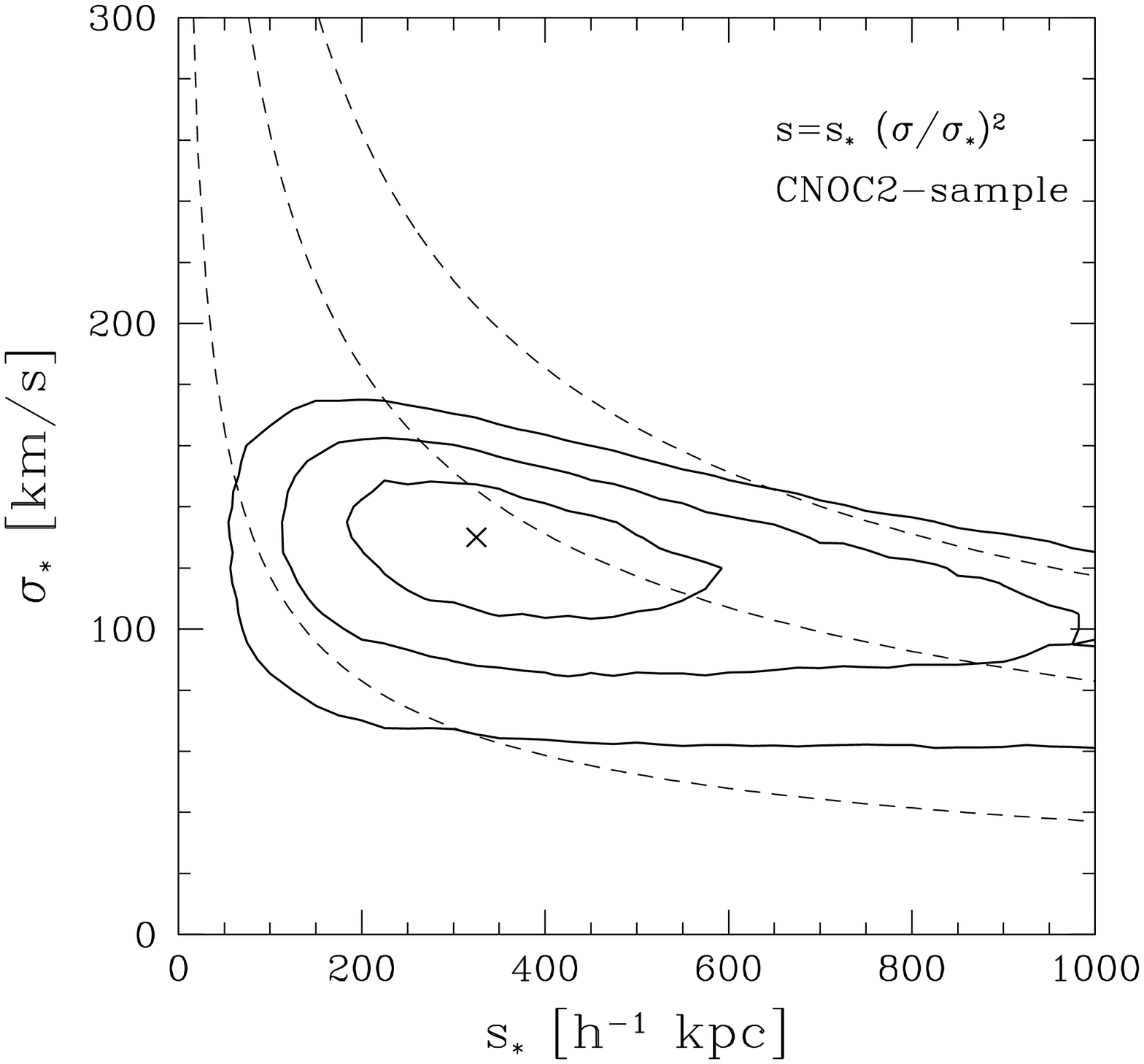}}
\begin{small}
\caption{Likelihood contours for the velocity dispersion $\sigma_*$ of
an $L^*_B(z=0)=5.6\times10^9~h^{-2} L_{B\odot}$ galaxy, and its
truncation radius $s_*$. The contours indicate 68.3\%, 95.4\%, and
99.7\% confidence limits on two parameters jointly. The cross
indicates the best parameters. We use $\sigma\propto L_B^{1/4}$, and
two different scaling relations for $s$. The dashed lines give models
with a total mass of $1,5,10\times10^{12} h^{-1}{\rm M}_\odot$.  (a)
The truncation radius $s$ is the same for all galaxies. (b) The
results when $s\propto \sigma^2$, which corresponds to the case where
all galaxy have the same total mass-to-light ratio.
\label{chi2gal}}
\end{small}
\end{center}
\end{figure*}

Figure~\ref{chi2gal} shows the results for the `CNOC2' sample, for
which the individual redshifts of the lenses are known. It shows the
likelihood contours for the parameters $\sigma_*$ and $s_*$ jointly.
The result for a constant $s$ for all galaxies is presented in
figure~\ref{chi2gal}a, and the result for $s\propto\sigma^2$ is shown
in figure~\ref{chi2gal}b. We omit the likelihood plot
$s\propto\sigma^4$, but we list the best fit parameters (68.3\%
confidence limits) for all three models in Table~\ref{trunctab}a.  The
results change slightly when we assume no luminosity evolution, and
the values of the parameters are listed in Table~\ref{trunctab}d. The
confidence intervals on $\sigma_*$ ($s_*$) listed in
Tables\ref{trunctab}a and d are obtained by marginalising over $s_*$
($\sigma_*$). Note that these confidence intervals cannot be inferred
directly from Figure~\ref{chi2gal} (e.g., Press et al. 1992, Numerical
Recipes in C, Figure 15.6.3).

We also examined the influence of an imperfect correction for PSF
anisotropy.  The change in the derived velocity dispersion is
approximately 1\% if we under or overcorrect by 10\%. The value of
$s_*$ is changed by $\sim 6\%$ in these case.  As we have demonstrated
above, the actual correction for PSF anisotropy is much better than
10\%, and hence we find it to be negligible.

In all cases the velocity dispersion is well constrained, and in good
agreement with the results from the ensemble averaged tangential
distortion. The value of the truncation parameter has significant
freedom, and the value of $s_*$ depends strongly on the assumed
scaling relation.  We find that the value of $s_*$ decreases when $s$
scales with a high power of the velocity dispersion. The minimum
$\chi^2$ for the three models are comparable, and no scaling
relation for $s$ is preferred over the other
ones. Table~\ref{trunctab}a also lists the best estimate and the
68.3\% confidence limits for the total mass and mass-to-light ratio
of an \lstar galaxy. The total galaxy mass is not well constrained,
mainly because of the large uncertainty in the value of the truncation
parameter $s_*$.

The mass model uses only half of the galaxies with $17.5<R<21.5$, and
therefore it ignores the contributions of the other galaxies (they are
given zero mass). If the remaining galaxies were distributed randomly,
their contribution to the lensing signal is that of noise.  We know,
however, that galaxies are clustered. Consequently the mass model will
associate the mass of the galaxies without redshifts with the galaxies
in the `CNOC2' sample that have measured redshifts.

\begin{table*}
\begin{center}
\begin{tabular}{lccccccc}
\hline
\hline
 & (1) & (2) & (3) & (4) & (5) & (6) & (7) \\
 & scaling $s$	& $\sigma_*$ & $s_*$          & $s_*^{\rm min}$ & $s_*^{\rm max}$ & ${\rm M}^*_{\rm tot}$ & ${\rm M}_{\rm tot}/{\rm L}^*_{\rm B}$ \\
 &		& [km/s]     & [$h^{-1}$ kpc] & [$h^{-1}$ kpc]  & [$h^{-1}$ kpc]  & $[10^{12} h^{-1}{\rm M}_\odot$] & [$h {\rm M}/{\rm L}_{{\rm B}\odot}$] \\
\hline
\hline
\multicolumn{8}{c}{luminosity evolution $\propto(1+z)$, ${\rm L}^*_{\rm B}(z=0)=5.6\times10^9~h^{-2} {\rm L}_{\rm {B}\odot}~$}\\
\hline
\hline
(a) & \multicolumn{7}{c}{galaxies with redshifts from CNOC2} \\
\hline
 & $\propto \sigma^0$  	& $126^{+14}_{-16}$	& $403^{+169}_{-99}$	 & 105 & 880	&  $4.7^{+1.8}_{-1.2}$ & $839^{+321}_{-214}$ \\
 & $\propto \sigma^2$   & $126^{+14}_{-16}$	& $313^{+136}_{-81}$ 	 & 86  & 772	&  $3.6^{+1.7}_{-0.8}$ & $643^{+304}_{-143}$ \\
 & $\propto \sigma^4$ 	& $124^{+14}_{-15}$	& $236^{+105}_{-52}$	 & 73  & 605	&  $2.6^{+1.5}_{-0.6}$ & $464^{+268}_{-107}$ \\
\hline
(b) & \multicolumn{7}{c}{`bright' sample, no individual redshifts used} \\
\hline
 & $\propto \sigma^0$  	& $110^{+15}_{-15}$	& $277^{+148}_{-112}$	 & 51  & 623 	& $2.4^{+1.3}_{-0.9}$ & $429^{+232}_{-161}$ \\
 & $\propto \sigma^2$  	& $111^{+15}_{-15}$	& $220^{+133}_{-100}$	 & 43  & 531	& $1.9^{+1.2}_{-0.8}$ & $339^{+214}_{-143}$ \\
 & $\propto \sigma^4$  	& $112^{+14}_{-14}$	& $163^{+103}_{-74}$	 & 34  & 407	& $1.4^{+0.9}_{-0.7}$ & $250^{+161}_{-125}$ \\
\hline
(c) & \multicolumn{7}{c}{`bright' sample, using individual redshifts when available} \\
\hline
 & $\propto \sigma^0$ 	& $110^{+12}_{-12}$	& $337^{+130}_{-100}$	 & 86  & 679	& $2.8^{+1.2}_{-0.8}$ & $500^{+214}_{-143}$ \\
 & $\propto \sigma^2$ 	& $111^{+12}_{-12}$	& $260^{+124}_{-73}$	 & 80  & 556	& $2.2^{+1.2}_{-0.7}$ & $393^{+214}_{-125}$ \\
 & $\propto \sigma^4$  	& $113^{+12}_{-12}$	& $195^{+92}_{-63}$	 & 68  & 432	& $1.6^{+1.0}_{-0.5}$ & $286^{+179}_{-89}$  \\
\hline
\hline
\multicolumn{8}{c}{no luminosity evolution, ${\rm L}^*_{\rm B}(z=0)=7.3\times10^9~h^{-2} {\rm L}_{\rm {B}\odot}~$}\\
\hline
\hline
(d) & \multicolumn{7}{c}{galaxies with redshifts from CNOC2} \\
\hline
 & $\propto \sigma^0$  	& $125^{+14}_{-16}$	& $432^{+181}_{-106}$	 & 112 & 942	&  $4.9^{+1.9}_{-1.3}$ & $678^{+260}_{-173}$ \\
 & $\propto \sigma^2$   & $125^{+14}_{-16}$	& $350^{+152}_{-90}$ 	 & 96  & 862	&  $4.0^{+1.9}_{-0.9}$ & $541^{+256}_{-120}$ \\
 & $\propto \sigma^4$ 	& $123^{+14}_{-15}$	& $267^{+119}_{-59}$	 & 82  & 684	&  $2.9^{+1.7}_{-0.7}$ & $396^{+228}_{-91}$ \\
\hline
(e) & \multicolumn{7}{c}{`bright' sample, no individual redshifts used} \\
\hline
 & $\propto \sigma^0$  	& $109^{+15}_{-15}$	& $297^{+159}_{-120}$	 & 55  & 727 	& $2.5^{+1.4}_{-0.9}$ & $346^{+261}_{-130}$ \\
 & $\propto \sigma^2$  	& $110^{+15}_{-15}$	& $246^{+149}_{-112}$	 & 48  & 593	& $2.1^{+1.3}_{-0.9}$ & $285^{+180}_{-120}$ \\
 & $\propto \sigma^4$  	& $111^{+14}_{-14}$	& $184^{+116}_{-84}$	 & 38  & 460	& $1.6^{+1.0}_{-0.8}$ & $213^{+137}_{-107}$ \\
\hline
(f) & \multicolumn{7}{c}{`bright' sample, using individual redshifts when available} \\
\hline
 & $\propto \sigma^0$ 	& $109^{+12}_{-12}$	& $361^{+139}_{-107}$	 & 92  & 727	& $2.9^{+1.3}_{-0.8}$ & $404^{+173}_{-115}$ \\
 & $\propto \sigma^2$ 	& $110^{+12}_{-12}$	& $290^{+139}_{-82}$	 & 89  & 621	& $2.4^{+1.3}_{-0.8}$ & $331^{+180}_{-105}$ \\
 & $\propto \sigma^4$  	& $112^{+12}_{-12}$	& $221^{+104}_{-71}$	 & 77  & 489	& $1.8^{+1.1}_{-0.6}$ & $244^{+153}_{-76}$ \\
\hline
\hline
\end{tabular}
\begin{small}
\caption{(a) Results from the maximum likelihood analysis using the
galaxies with redshifts from the CNOC2 survey. (b) The results for the
analysis of the `bright' sample. (c) Results when we use the observed
redshifts if available, and the redshift distribution if the redshift
has not been measured. (d)-(f) list the results for no luminosity
evolution, for which case ${\rm L}^*_{\rm B}(z=0)=7.3\times10^9~h^{-2}
{\rm L}_{\rm {B}\odot}~$. Only the results presented in (a) and (d)
are real maximum likelihood parameters. See the text for more details.
The errors correspond to 68.3\% confidence.  (1) the scaling relation
used for the truncation parameter $s$; (2) estimate for $\sigma_*$;
(3) estimate for $s_*$; (4) 99.7\% confidence lower limit on $s_*$;
(5) 95\% confidence upper limit on $s_*$; (6) estimate for the total
mass; (7) total mass-to-light ratio of an \lstar\ galaxy.
\label{trunctab}}
\end{small}
\end{center}
\end{table*}  

To examine this in more detail we study the `bright' sample, which is
selected irrespective of redshift information. Even this selection is
not ideal, neighbouring galaxies that are fainter than the applied
magnitude limits are still excluded. If a substantial fraction of the
mass is in these galaxies both the values of $\sigma_*$ and $s_*$ are
overestimated.

\begin{figure}
\begin{center}
\leavevmode
\hbox{%
\epsfxsize=8cm
\epsffile{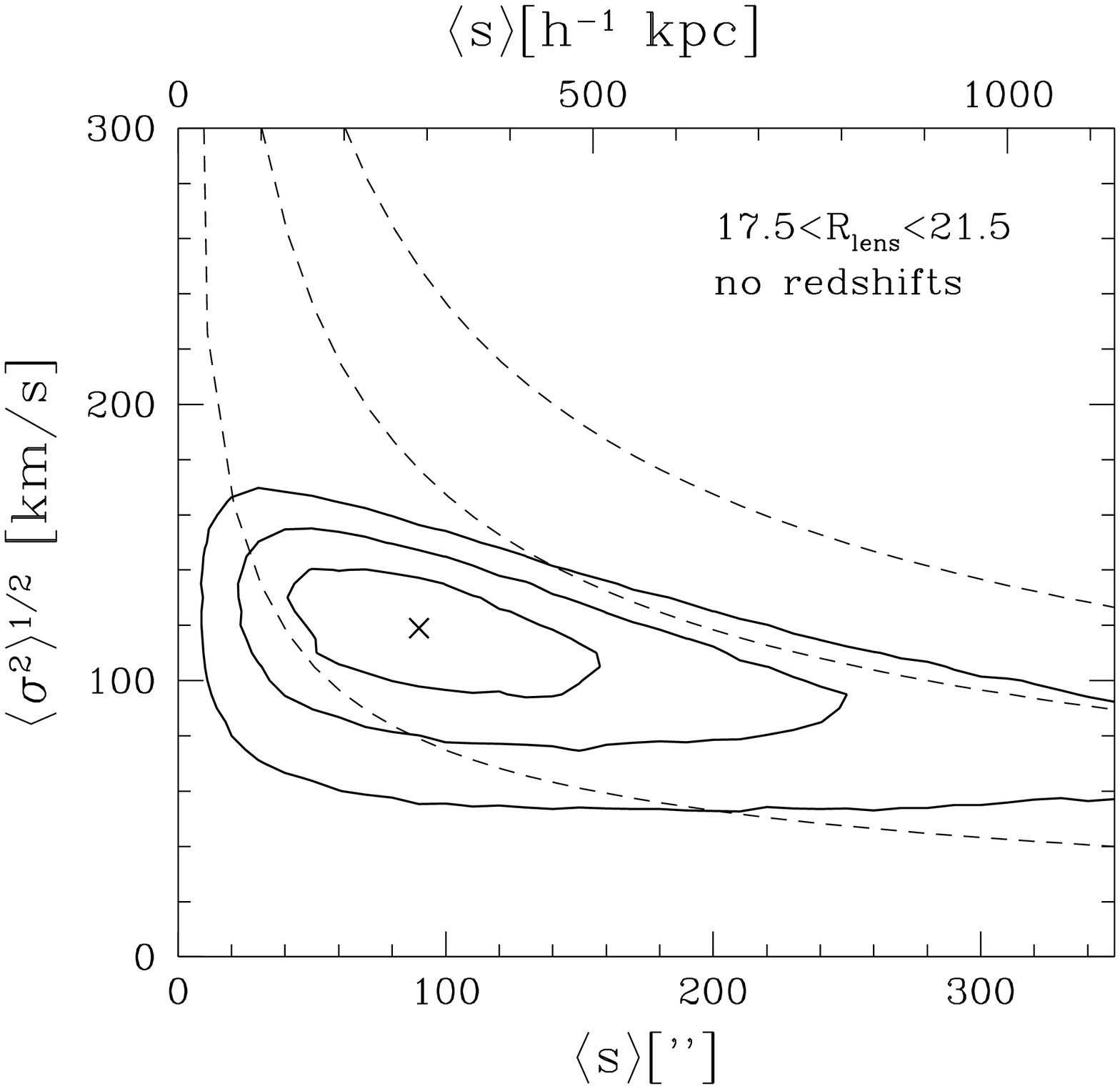}}
\begin{small}
\caption{Likelihood contours for the velocity dispersion $\sigma_*$ of
an $L^*_B(z=0)=5.6\times10^9~h^{-2} L_{B\odot}$ galaxy, and the
average value of the truncation radius $\langle s \rangle$ a
determined from the `bright' sample.  We have also indicated the
physical scale when $s$ is the same for all galaxies. The contours
indicate 68.3\%, 95.4\%, and 99.7\% confidence limits on two
parameters jointly. The cross indicates the best parameters. The
dashed lines give models with a total mass of $1,5,10\times10^{12}
h^{-1}{\rm M}_\odot$.
\label{chi2_noz}}
\end{small}
\end{center}
\end{figure}

Figure~\ref{chi2_noz} shows the resulting likelihood contours. To
obtain figure~\ref{chi2_noz} we have assumed that all halos have the
same velocity dispersion, and the same value for $s$ in arcseconds.
We find $\langle\sigma^2\rangle^{1/2}=117^{+13}_{-15}$ km/s, and
$\langle s \rangle=88^{+43}_{-27}$ arcseconds (68.3\% confidence). At
the average redshift of the lenses, this corresponds to $\langle s
\rangle=284^{+139}_{-87} h^{-1}{\rm kpc}$. The estimate for $\langle
s\rangle$ is a weighted average over the value of $s$ for all the
galaxies, which are all at different redshifts. The interpretation of
the result is not straightforward for the various options for the
scaling relation of $s$.

To infer the best estimates for $\sigma_*$, and $s_*$ of the bright
sample one has to perform a maximum likelihood analysis in which the
redshift of each individual galaxy is a free parameter, which has to
be chosen such that it maximizes the likelihood. This approach
is computationally not feasible, and we use another approach to
obtain estimates for $\sigma_*$ and $s_*$.

We use the same method that was used for the galaxies with redshifts,
but instead we create mock redshift catalogs. The redshift distribution
of the galaxies in the 'bright' sample has been measured by the
CNOC2 survey. The redshifts of the galaxies in the mock catalogs
are drawn randomly for the CNOC2 survey based on the apparent $R$
magnitude of the galaxies in the 'bright' sample. We take the
incompleteness of the survey into account when the redshifts are drawn
from the survey.  This mock redshift catalog allows us to determine
the best estimates for $\sigma_*$ and $s_*$, based on a maximum
likelihood analysis of the observed distortion field. We repeat this
procedure 25 times, and use the average $\chi^2$ surface. The best
estimates for $\sigma_*$ and $s_*$ for the three different scaling
relations for $s$ are listed in Table~\ref{trunctab}b.

We verified that this procedure yields an unbiased estimate, by
simulating the data and applying the procedure described above.  We
used the derived values of $\sigma_*$ and $s_*$ (see
Table~\ref{trunctab}b) as input values for the simulations.  These
simulations also allow us to estimate the scatter in the
measurements. Compared to the situation where the redshift of each
lens is known, the lack of redshifts increases the uncertainty in
$s_*$ by $\sim 50\%$. The uncertainty in $\sigma_*$ is increased by
30\%.  For instance, adopting $s\propto\sigma^2$, the uncertainty
increases by $\sim 45h^{-1}$ kpc and the uncertainty in $\sigma_*$
increases by $\sim 3$ km/s.

Comparison of the results presented in Table~\ref{trunctab}b with the
results obtained for the `CNOC2' sample (Table~\ref{trunctab}a) shows
that the estimates for $\sigma_*$ and $s_*$ are smaller, but
consistent with one another. For the case that $s\propto\sigma^2$ the
'bright' sample yields $\sigma_*=111\pm15$ km/s and
$s_*=220^{+133}_{-100} h^{-1}$ kpc, compared to
$\sigma_*=126^{+15}_{-16}$ km/s and $s_*=313^{+136}_{-81} h^{-1}$ kpc
for the 'CNOC2' sample. The agreement with the results from the direct
averaging method is good.

The analysis of the `bright' sample has ignored any available redshift
information. To make optimal use of the available information, we redo
the analysis as follows. For the galaxies with redshifts we use the
observed values, but for the remaining galaxies we draw redshifts and
luminosities from the CNOC2 survey randomly (as was done previously
for all galaxies in the `bright' sample). As before, we create mock
catalogs which are analysed. This procedure improves the accuracy in
the estimates of $\sigma_*$ and $s_*$. The results are presented in
Table~\ref{trunctab}c. Simulations show that the lack of redshifts for
all galaxies with $19.5<R<21$ from CNOC2 increases the uncertainty in
$\sigma_*$ by 18\% and the uncertainty in $s_*$ by 10\%, compared to
the situation where full redshift information is available.

\subsection{Effect of galaxy groups}

It is well known that galaxies cluster, and most of the galaxies
reside in groups of galaxies. If the matter in galaxy groups is
associated with the halos of the group members (i.e. these halos are
indistinguishable from the halos of isolated galaxies) the analysis
presented above gives a fair estimate of the sizes of galaxy
halos. However, if a significant fraction of the dark matter in galaxy
groups is distributed in a common group halo, the interpretation of
the results becomes difficult.

Fischer et al. (2000) argued that galaxy groups complicate the
interpretation of the galaxy-galaxy lensing signal.  They measured the
lensing signal out to large projected distances, and find that
the distortion does not vanish at large radii, which is 
caused by large scale clustering (the angular correlation
function declines relatively slowly).

Compared to Fischer et al. (2000) we have the advantage that a number
of galaxy groups have been identified in the CNOC2 fields (Carlberg et
al.  2001). The weak lensing signal of these groups was studied by
Hoekstra et al. (2001a). They derived a mass weighted group velocity
dispersion of $\langle\sigma^2_{\rm group}\rangle^{1/2}=273^{+48}_{-59}$ 
km/s.

We will use the groups to examine their effect on the galaxy-galaxy
lensing results for the galaxies with redshifts from the CNOC2 survey. 
In addition to the galaxies, we include the lensing signal from 
group halos to our mass model. The groups are placed at their observed
positions, and the groups are modeled as singular isothermal spheres.
To study the effect of the groups on the galaxy-galaxy lensing results
we increase the velocity dispersion of the `groups' from 0 km/s (the
result without groups) up to 300 km/s. For each choice of the 
group velocity dispersion we perform a maximum likelihood analysis and
determine the best estimates for $\sigma_*$ and $s_*$. 

For $\sigma_{\rm group}<150$ km/s the best estimates for $\sigma_*$
and $s_*$ vary by a few percent. For larger group velocity dispersions
the value for $\sigma_*$ and $s_*$ decrease slowly with increasing
$\sigma_{\rm group}$. For a group velocity dispersion of 300 km/s,
the value of $\sigma_*$ has dropped by $\sim 10\%$, and the
value of $s_*$ has decreased by $\sim 20\%$. We note that the
minimum $\chi^2$ has increased significantly for $\sigma_{\rm group}=300$ km/s,
compared to $\sigma_{\rm group}=0$ km/s. 

Ideally one would like to use different halo models for the group
members and the `isolated' galaxies and study the difference in the
best parameters for these two types of galaxies. With the current data
we cannot perform such an analysis, because the number of group
members is too low. However, our results are based on approximately
one quarter of the full CNOC2 data set. An analysis of the full survey
will improve the signal-to-noise ratio of the measurements by a factor
$\sim 2$, and will allow a better study of the effect of galaxy
groups. Also numerical simulations can be useful to examine if, and
how one can separate the contribution of the galaxy halos and the
smooth group halos.

\subsection{Comparison with other lensing studies}

Other studies selected different samples of lenses, used different
scaling relations, or made other assumptions about the luminisoty
evolution. Fortunately, sufficient information is available to allow
for a useful, if crude, comparison. We will compare the findings of
some other studies to the results from our maximum likelihood
analysis.

The typical lens galaxy in the sample studied by Hudson et al. (1998)
is at $z\sim 0.6$, and has $M_B=-18.5+5\log h$ (for $q_0=0.5$).
Hudson et al. (1998) derived a circular velocity of $210\pm40$ km/s.
This corresponds to a galaxy with a luminosity of $L_B(z=0)\sim
3.2\times 10^9~h^{-2} L_{B\odot}$ for our choice of cosmology and
luminosity evolution. Our weak lensing analysis suggests a circular
velocity of $V_c=119\pm12$ km/s for such a galaxy. If we assume no
luminosity evolution (as Hudson et al. (1998) did), we obtain a
circular velocity of $V_c=130\pm14$. The two results are inconsistent
at the $\sim 2\sigma$ level. It is not clear, however, how to
interpret the difference in circular velocity, as our weak lensing analysis 
probes a larger physical scale than Hudson et al. (1998), and the 
mix of galaxy types might also be different. 

Fischer et al. (2000) measured the galaxy-galaxy lensing signal with
much higher accuracy than Hudson et al. (1998) from SDSS data. Fischer
et al. (2000) find for their sample of lenses an average mass weighted
velocity dispersion of $\langle \sigma^2\rangle^{1/2}=145-195$ km/s
(95\% confidence). McKay et al. (2001) noted that the SDSS redshift
survey showed that the lens redshifts used by Fischer et al. (2000)
were overestimated by $\sim 35\%$. The correct range in line-of-sight
velocity dispersion of the Fischer et al. (2000) analysis should be
$\langle \sigma^2\rangle^{1/2}\approx105-145$ km/s.

Most of the signal comes from galaxies with luminosities around
$L_{g'}=8.7\times 10^9 h^{-2} {\rm L}_{g'\odot}$ (P.~Fischer, private
communication). If we use an average $B-V=0.55$ and the
transformations from Fukugita et al. (1996), we find that the adopted
$B$ band luminosity of our \lstar\ galaxy corresponds to a $g'$ band
luminosity of $L_{g'}=6\times 10^9 h^{-2} {\rm L}_{g'\odot}$.  Our
lensing analysis implies a velocity dispersion of
$\sigma=122^{+13}_{-12}$ km/s (68.3\% confidence) for a galaxy with
$L_{g'}=8.7\times 10^9 h^{-2} L_{g'\odot}$, which agrees well
with the Fischer et al. (2000) result.

McKay et al. (2001) used a sample of galaxies with spectroscopic
redshifts from the SDSS. Their sample of lenses is comparable in size
to the one used by Fischer et al. (2000). Using the complete sample,
McKay et al. (2001) find a velocity dispersion $\sigma=100-130$ km/s
(95\% confidence) with a best fit value of $\sigma=113$ km/s for a
galaxy with $L_{g'\odot}\sim 9\times 10^9 h^{-2} {\rm L}_{g'\odot}$.
For such a galaxy, our lensing analysis yields
$\sigma=122^{+13}_{-12}$ km/s (68.3\% confidence), in excellent
agreement with McKay et al. (2001).

Fischer et al. (2000) and McKay et al. (2001) also attempted to
constrain the sizes of galaxy halos. Both studies indicate that the
the galaxy halos are large. Unfortunately it is not clear what the
correct value for the Fischer et al. (2000) result is. McKay et
al. (2001) used the same approach as Fischer et al. (2000) and find a
lower limit of $s^{\rm min}=230h^{-1}$ kpc (95\% confidence). McKay et
al. (2001) do not use any scaling for $s$ (they assume all halos are
the same) and their results should be compared to our results for
constant $s$ for all galaxies. We obtain an estimate of
$s=s_*=337^{+130}_{-100}h^{-1}$ kpc, which is in good agreement with
McKay et al. (2001).

Interestingly, McKay et al. (2001) find that the mass within an
aperture of radius $260h^{-1}$ kpc scales proportional to the
luminosity of the lens. This appears to be in contradiction with the
naive expectation from the Tully-Fischer or Faber-Jackson relation,
which suggests $M\propto\sqrt L$. However, the latter is no longer
true for truncated halos. As discussed in Section~5.3, the relation
between the mass and luminosity on large scales depends on the adopted
scaling relation for $s$ (at large radii $M\propto \sigma^2s$).  For
instance, $s\propto\sigma^2$ results in $M \propto L$ at radii (much)
larger than $s$.

Hence, the results presented by McKay et al. (2001) suggest that the
truncation is smaller than $260h^{-1}$ kpc. This is likely to be the
case for the faint lenses, but the brighter lenses are expected to
have larger halos. However, luminous galaxies are clustered more strongly
(e.g., Norberg et al. 2002) and the contribution from neighboring
galaxies could bias the lensing signal somewhat high, although
the effect is expected to be small (McKay et al. 2001).

Recently, Yang et al. (2002) used numerical simulations to study the
results presented by McKay et al. (2001). The galaxies in the
simulations obey the Tully-Fischer or Faber-Jackson relation.  Yang et
al. (2002) find that the aperture mass within $260h^{-1}$ kpc is
proportional to the luminosity of the lens.  The results from Yang et
al. (2002), however, suggest that the linear relation found by McKay
et al. (2001) is a coincidence, because they find that the halo mass
scales $\propto L^{1.5}$ (which corresponds to
$s\propto\sigma^4$). The mass with an aperture of radius $260h^{-1}$
kpc results in a linear relation, whereas other apertures would yield
different results.

Finally we note that the results presented here are in good agreement
with preliminary results from an analysis of $R$ band imaging data
from the Red-Sequence Cluster Survey (RCS). Hoekstra, Yee, \& Gladders
(2001d) use lenses with $19.5<R<21$. Their results imply $\sigma_*=102
\pm 4$ km/s and $s_*=198^{+25}_{-19} h^{-1}$ kpc for an \lumstar\
galaxy (adopting $s\propto\sigma^2$), which is in good agreement with
our results.

\subsection{Comparison with Tully-Fisher relation}

Comparison of the luminosity and the rotation velocity of gas in
spiral galaxies has shown that there exists a tight relation between
these two observables: the Tully-Fisher (TF) relation. Note, however,
that weak lensing probes the mass on a much larger scale than the
rotation curve.  For example, if the rotation curve is slowly
declining, the lensing result will be lower (compared to the flat
rotation curve). Furthermore, the sample of lenses used hereis
different from the sample of spiral galaxies used for the TF
relation. Ideally one would like to select a sample of spiral galaxies
for this comparison, but this is outside the scope of this paper.
Nevertheless, it is interesting to compare our lensing
result to a local determination of the TF relation.

One of the most recent determinations of the TF relation was presented
by Verheijen (2001), who studied a large sample of galaxies in the
Ursa Major cluster of galaxies. The sample has the advantage that the
galaxies are at a distance of 18.6 Mpc (Tully \& Pierce 2000). This
allowed Verheijen (2001) to study the scatter in the TF relation.

\begin{figure}
\begin{center}
\leavevmode
\hbox{%
\epsfxsize=8cm
\epsffile{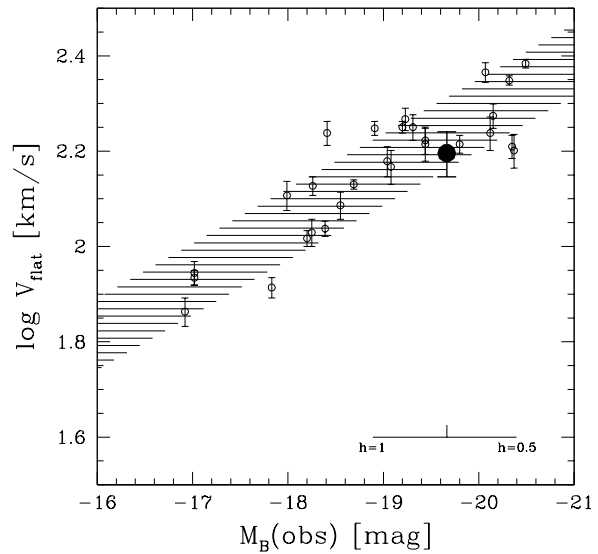}}
\begin{small}
\caption{Plot of rotation velocity of the `flat' part of the rotation
curve $V_{\rm flat}$ versus the absolute B-band magnitude (uncorrected
for inclination and extinction). The open dots with error bars
indicate the measurements of Verheijen (2001), using the (uncorrected)
magnitudes from Tully et al. (1996).  The hatched region indicates the
$1\sigma$ interval around the derived TF relation.  In the comparison
we use the circular velocity corresponding to the value of $\sigma_*$
(assuming $s\propto \sigma^2$) listed in Table~\ref{trunctab}c. The
result is indicated by the large black dot.  The absolute magnitudes
of the galaxies from Verheijen (2001) are based on the adopted
distance to the Ursa Major cluster, and therefore do not depend on the
value of the Hubble parameter. The inferred luminosities of the
galaxies studied in the weak lensing analysis do depend on $H_0$, and
for the comparison we take a value of $H_0=70$ km/s/Mpc.  We have also
indicated how the weak lensing result would change with $h$.  The weak
lensing result is in good agreement with the observed TF relation.  We
note that weak lensing probes the mass on a much larger scale than the
rotation curve, and one therefore might expect only a qualitative
agreement. Furthermore, the sample of lenses is rather different from
the sample of spiral galaxies used for the TF relation.
\label{comp_tf}}
\end{small}
\end{center}
\end{figure}

Verheijen (2001) has measured $V_{\rm flat}$, the rotation velocity of
the `flat' part of the rotation curve, which is representative for
the mass of the galaxy. As is customary in TF studies, Verheijen
(2001) used extinction corrected luminosities. The luminosities used
in our lensing analyses, however, have not been corrected for
inclination and extinction. To allow a direct comparison with the weak
lensing results, we use the uncorrected magnitudes from Tully et
al. (1996). The resulting TF relation is presented in
Figure~\ref{comp_tf} (open dots with error bars).

The hatched region indicates the $1\sigma$ interval around the TF
relation, based on the scatter in the observations. In the comparison
we use the circular velocity corresponding to the value of $\sigma_*$
(assuming $s\propto \sigma^2$) listed in Table~\ref{trunctab}c. We
assume that $V_{\rm flat}=\sqrt{2}\sigma$. The result is indicated by
the large black dot. The comparison depends on the value of the Hubble
parameter, because the inferred luminosities of the lenses in our
analysis depend on $h$, whereas the absolute magnitudes of the
galaxies from Verheijen (2001) are based on the adopted distance to
the Ursa Major cluster (and therefore do not depend on the Hubble
parameter). For the comparison presented in Figure~\ref{comp_tf} we
adopted a value of $H_0=70$ km/s/Mpc. Although one cannot make a
direct comparison between the weak lensing results and the TF relation
(for reasons mentioned above) it is comforting that the results are
rather similar.

\section{Mass-to-light ratio and $\Omega_m$}

A useful method to estimate the matter density of the universe was
proposed by Oort (1958): $\Omega_m$ is the product of the average
mass-to-light ratio of the universe, and its luminosity
density. Carlberg et al. (1997) measured the mass-to-light ratios of
rich clusters of galaxies, and inferred $\Omega_m=0.19\pm0.06$.  The
galaxy properties of rich clusters are different from those of the
field, and a large correction is needed to relate the cluster
mass-to-light ratio to that of the field. Smaller corrections are
required when galaxy groups are used. Hoekstra et al. (2001a) derived
$\Omega_m=0.19\pm 0.1$ from their weak lensing analysis of 50
groups. Bahcall \& Comerford (2002) found $\Omega_m=0.17\pm0.05$ from
a combination of clusters and groups.

A measurement of the average mass-to-light ratio of field galaxies has
the advantage that no additional corrections for the difference in
stellar populations are required. An estimate for the average
mass-to-light ratio of the field, i.e., the universe as a whole, can
be obtained from the results of our maximum likelihood analysis The
inferred total mass-to-light ratio, however, depends on the assumed
scaling relation for the truncation parameter $s$ (as is apparent from
column~7 in Table~\ref{trunctab}).

Lin et al. (1999) have determined the luminosity function of field
galaxies from the CNOC2 survey, and we use their results to estimate
the average mass-to-light ratio of the field. The results are listed
in Table~\ref{omegatab} (column~4). When $s\propto\sigma^2$, the
average mass-to-light ratio equals that of an \lstar\ galaxy.  In the
case that all halos have the same value for $s$, we already found that
the resulting mass-to-light ratio of an $L_*$ galaxy is high, but in
addition, fainter galaxies have even higher mass-to-light ratios.
This results in a very high field mass-to-light ratio. On the other
hand, if $s\propto \sigma^4$, faint galaxies have low mass-to-light
ratios, and the field mass-to-light ratio is somewhat lower than that
of an \lstar\ galaxy.

\begin{table*}
\begin{center}
\begin{tabular}{cccccc}
\hline
\hline
(1) & (2) & (3) & (4) & (5) & (6) \\
scaling $s$ & scaling M/L$_B$   & ${\rm M}_{\rm tot}/{\rm L}^*_{\rm B}(z=0)$ & $\langle{\rm M}/{\rm L}_{\rm B}\rangle(z=0)$  & $\langle{\rm M}/{\rm L}_{\rm B}\rangle(z=0.34)$ & $\Omega_m$\\
	    &	    		& [$h {\rm M}/{\rm L}_{{\rm B}\odot}$]  & [$h {\rm M}/{\rm L}_{{\rm B}\odot}$]     &	[$h {\rm M}/{\rm L}_{{\rm B}\odot}$] &	\\
\hline
$\propto \sigma^0$ & $\propto 1/\sqrt{L_B}$	& $500^{+214}_{-143}$ & $850^{+364}_{-243}$ & $634^{+272}_{-181}$ & $0.69^{+0.30}_{-0.20}$	 \\
$\propto \sigma^2$ & constant			& $393^{+214}_{-125}$ & $393^{+214}_{-125}$ & $293^{+160}_{-93}$  & $0.32^{+0.17}_{-0.10}$	 \\
$\propto \sigma^4$ & $\propto \sqrt{L_B}$	& $286^{+179}_{-89}$  & $240^{+150}_{-75}$  & $168^{+112}_{-56}$  & $0.18^{+0.12}_{-0.06}$	 \\
\hline
\hline
\end{tabular}
\begin{small}
\caption{Estimates for the average field mass-to-light ratio for
different scaling relations of the truncation parameter $s$.  The
errors correspond to 68.3\% confidence.  (1) the scaling relation used
for the truncation parameter $s$; (2) the dependence of the total
mass-to-light ratio on the luminosity; (3) total mass-to-light ratio
of an \lstar\ galaxy; (4) average field mass-to-light ratio at $z=0$;
(5) average field mass-to-light ratio at the average redshift of the
sample of galaxies $(z=0.34)$; (6) estimate for the matter density of
the universe $\Omega_m$.
\label{omegatab}}
\end{small}
\end{center}
\end{table*}  

To estimate the luminosity density of the universe we use the results
from Lin et al. (1999), which are also based on the CNOC2 survey.  We
convolve the redshift distribution of the galaxies with the redshift
dependent luminosity density from Lin et al. (1999), which yields
$j=(3.0\pm0.6) \times10^8 h {\rm L}_{{\rm B}\odot}{\rm Mpc}^{-3}$.
We use this result, and our estimates for the average field
mass-to-light ratio at $z=0.34$ to derive the corresponding values
of $\Omega_m$, which are listed in column~6 of Table~\ref{omegatab}. 

With the current data, the statistical error in the value of
$\Omega_m$ is large. In addition, the uncertainty in the scaling
relation of $s$ limits the determination of $\Omega_m$ through this
technique. We also assumed that all matter in the universe is
associated with the galaxy dark matter halos. The lensing signal is
not changed by adding a sheet of constant surface density (Gorenstein,
Shapiro, \& Falco 1988). Thus a uniformly distributed form of dark
matter cannot be detected in our analysis. Consequently, our estimate
for $\Omega_m$ should be interpreted as a lower limit.  Another
approach is to use independent measurements of $\Omega_m$ (e.g., from
CMB or cosmic shear studies) to constrain the scaling relation for
$s$.

\section{Conclusions}

We observed two blank fields of approximately 30 by 23 arcminutes
using the William Herschel Telescope. The fields were studied as part
of the Canadian Network for Observational Cosmology Field Galaxy
Redshift Survey (CNOC2) (e.g., Yee et al. 2000; Carlberg et al. 2001),
and spectroscopic redshifts are available for 1125 galaxies in the two
fields. Earlier results on groups of galaxies, based on these data,
were presented by Hoekstra et al. (2000a).

We have measured the lensing signal caused by large scale structure.
We observed signal is low, and consistent with more accurate
measurements from much larger surveys (e.g., Bacon et al. 2002;
Hoekstra et al. 2002a,b; Refregier et al. 2002; van Waerbeke et
al. 2002). 

We examined the effect of an imperfect correction for PSF anisotropy,
and find the correction used here has worked well: it gives the
smallest cosmic shear signal. In addition we find that the uncertainty
in the correction has a negligible effect on the galaxy-galaxy lensing
results.

We have studied the ensemble averaged tangential distortion
(galaxy-mass correlation function) around three subsamples of
lenses. In all three cases a clear lensing signal is detected. We
relate the lensing signal to an estimate of the velocity dispersion of
an \lstar\ galaxy, and find that the results for the three samples
agree well with each other. For the sample of galaxies with redshifts
from the CNOC2 survey we derive $\sigma_*=130^{+15}_{-17}$ km/s (or
$V_c^*=184^{+22}_{-25}$ km/s). Note, however, that this measurement
is slightly biased , because of the clustering of the lenses.

To study the properties of the dark matter halos surrounding the lens
galaxies, we used a maximum likelihood analysis. This technique
allowed us to derive constraints on both the velocity dispersion and
the extent of the dark matter halos (it naturally accounts for the
clustering of the lenses). The value of the truncation parameter $s$
depends strongly on the assumed scaling relation, but the velocity
dispersion is well constrained.  Galaxy groups can complicate the
interpretation of the truncation parameter $s$.  We examined how
galaxy groups affect our results, and we find that their effect is
relatively small. Our data are not sufficient for a detailed study,
but the full CNOC2 data set can be useful to this end, as well as
studies of numerical simulations.

Under the assumption that all galaxies have the same total
mass-to-light ratio, we find a value of $\sigma_*=111\pm12$ km/s (68\%
confidence, marginalised over the truncation parameter $s$) for the
velocity dispersion of an \lstar\ galaxy. For the truncation parameter
we obtain $s_*=260^{+124}_{-73}~h^{-1}$ kpc (68\% confidence,
marginalised over $\sigma_*$), with a 99.7\% confidence lower limit of
$80~h^{-1}$ kpc, and a 95\% confidence upper limit of $556~h^{-1}$
kpc. For this model we find that the average field mass-to-light ratio
at $z=0$ is $393^{+214}_{-124}$ \moverl, comparable to what is found
for rich clusters of galaxies (Carlberg et al.  1997) and galaxy
groups (Hoekstra et al. 2001a).

The field mass-to-light ratio provides an estimate of $\Omega_m$, the
matter density of the universe. The results from the weak lensing
analysis provide a lower limit on $\Omega_m$, because lensing cannot
detect dark matter that is distributed uniformly through the
universe. Unfortunately, our estimate of the field mass-to-light
ratio, and consequently the derived value of $\Omega_m$, depends
strongly on the assumed scaling of $s$.

The results presented here demonstrate that weak lensing is a powerful
tool to study the dark matter halos of field galaxies, as it can probe
the mass distribution out to large projected distances. In this paper
we have used approximately one quarter of the CNOC2 survey. An
analysis of the full survey will improve the accuracy of the results
by a factor $\sim 2$. Furthermore, many large imaging surveys are
currently underway. First results from the Red-Sequence Cluster Survey
(RCS) indicate that these surveys can provide accurate constraints on
the dark matter halos of field galaxies (e.g., Hoekstra et al. 2002d).

\section*{Acknowledgements}

It is a pleasure to thank the members of the Canadian Network for Observational
Cosmology without whom this project would never have been possible. We thank 
Jocelyn B{\'e}zecourt for his help to obtain part of the imaging data. 
The WHT observations for this project have been supported financially by the 
European Commission through the TMR program  'Access to large-scale 
facilities', awarded to the Instituto de Astroficica de Canarias. HH 
acknowledges support from the Leidsch Kerkhoven-Bosscha Fonds and the 
University of Toronto.

\end{document}